\documentclass[12pt]{elsart}
\usepackage{graphicx,amsmath}
\newcommand{\lsim}{\stackrel{\scriptstyle <}{\phantom{}_{\sim}}}
\newcommand{\gsim}{\stackrel{\scriptstyle >}{\phantom{}_{\sim}}}
\begin{document}

\begin{frontmatter}
	\title{Cooling of neutron stars in ``nuclear medium cooling scenario" with stiff equation of state including hyperons
	}
	\author[LIT,YER]{H.~Grigorian},
	\author[MEPHI,BLTP]{D.N.~Voskresensky} \and
	\author[MEPHI,BLTP]{K.A.~Maslov}
	\address[LIT]{Laboratory for Information Technologies, Joint Institute for Nuclear Research, RU-141980 Dubna, Russia}
	\address[YER]{Yerevan State University, Alek Manyukyan 1, 0025 Yerevan, Armenia}
	\address[MEPHI]{National Research Nuclear  University (MEPhI), RU-115409 Moscow, Russia}		\address[BLTP]{Bogoliubov Laboratory for Theoretical Physics, Joint Institute
			for Nuclear Research, RU-141980 Dubna, Russia}
		
	\begin{abstract}
		We demonstrate that the existing neutron-star cooling data  can be
		appropriately described within ``the nuclear medium cooling scenario" including hyperons under the assumption that different sources have different masses. We use a stiff equation of state of the relativistic mean-field model MKVORH$\phi$ with hadron effective couplings and masses dependent on the scalar field. It fulfills a large number of experimental constraints on the equation of state of the nuclear matter including  the $2\,M_{\odot}$ lower bound for the maximum predicted neutron-star mass and the constraint for the pressure from the heavy-ion particle flow.  We select appropriate $^1S_0$  proton and $\Lambda$ hyperon pairing gap profiles from those exploited in the literature and allow for a variation of the effective pion gap controlling the efficiency of the medium modified Urca process. The $^3P_2$ neutron pairing gap is assumed to be negligibly small in our scenario. The possibility of the pion, kaon and charged $\rho$-meson condensations is for simplicity suppressed. The resulting cooling curves prove to be sensitive to the value and the density dependence of the pp pairing gap and rather insensitive to the values of the $^1S_0$ neutron pairing gaps.
	\end{abstract}
\end{frontmatter}

\tableofcontents

\section{ Introduction}

 During many years the problem of the cooling of neutron stars (NSs) attracted great interest, e.g. see \cite{Bahcall:1965zzb,Tsuruta:1979,ST83,Migdal:1990vm,Yakovlev:2000jp,Voskresensky:2001fd,Page:2004fy,
	Potekhin:2015qsa,Schmitt:2017efp,Sedrakian:2018ydt}.
After first tens of seconds, or at most several hours for most massive NSs, the typical temperature of a newly born NS decreases below the neutrino-opacity temperature $T_{\rm opac}\sim $~1~MeV \cite{ST83,Sawyer:1978qe,Voskresensky:1986af,Migdal:1990vm,Voskresensky:2001fd}. Below we will be interested in the stage when
$T<T_{\rm opac}$. At this stage that lasts for the first $\sim 10^5$ yr. NSs are cooled by the direct neutrino radiation from  their interiors and then, for $t\gsim 10^6$ yr., by the photon radiation from the surface. For $T<T_{\rm opac}$ the typical averaged neutrino energy is $\sim$ several $T$, which is much larger than the nucleon particle width $\Gamma_N \sim T^2/\varepsilon_{{\rm F}}$.  Therefore nucleons can be treated within the quasiparticle approximation~\cite{Knoll:1995gs,Knoll:1995nz,Voskresensky:2001fd}, and the efficiency of processes of neutrino production can be graded simply according to their available phase space.

The direct Urca (DU) one-particle reactions, e.g. $n\to pe\bar{\nu}$, yield the largest neutrino emissivity, $\epsilon_{\nu}^{\rm DU}\sim 10^{27} T_9^6\Theta(n-n_{c,N}^{\rm DU})$ $\frac{{\rm erg}}{{\rm s}\cdot {\rm cm}^3}$ for non-superfluid systems, $T_9 =T/10^9$ K, $\Theta(x)$ is the step-function,  cf. \cite{Lattimer:1991ib}.
 Thereby the DU processes on nucleons exist  only in NSs with masses $M>M_{c}^{\rm DU}$ in which the central density exceeds the value $n_{c,N}^{\rm DU}$. An estimate of the DU-threshold follows from the triangle inequality for momentum conservation, that demands  the proton concentration in the $npe\mu$ matter  to be above 11.1-14.8\%. Therefore the DU processes set in at very different critical
densities depending on the density dependence of the symmetry energy, which is  an important quantity for the description of both heavy-ion reactions and  NSs. The DU emissivity proves to be so high that, if DU processes occurred in majority of NSs, the latter could not be visible in soft $X$-rays. It should, however, not be expected that the
objects observed in X-rays are some exotic family of NSs
rather than typical NSs. As a "weak DU constraint"  for an equation of state (EoS) \cite{Klahn:2006ir} suggested to require that $M_{c}^{\rm DU}>1.35 \, M_{\odot}$ since measured masses of the majority of NSs  in binaries lie between  $(1.35 -1.4) \, M_{\odot}$.
On the other hand, according to the pulsar population modeling most of NSs have masses below $1.5\,M_\odot$, see \cite{Alsing:2017bbc},  Figs. 1, 2 there.  Thereby \cite{Kolomeitsev:2004ff,Blaschke:2004vq,Klahn:2006ir} suggested to require the absence of the DU processes in NSs with $M<1.5 \, M_{\odot}$.  Reference \cite{Klahn:2006ir} named the requirement $M_{c}^{\rm DU}>1.5 \, M_{\odot}$ a ``strong DU constraint". The DU constraint puts a restriction on the density dependence of the symmetry energy.

The modified Urca (MU) two-nucleon processes have a much smaller emissivity than DU. Among them the luminosity of the neutron branch $nn\to npe\bar{\nu}$ of MU process is greater than that of the proton branch $np\to ppe\bar{\nu}$ and the nucleon bremstrahlung (NB) $nn\to nn\nu\bar{\nu}$, $np\to np\nu\bar{\nu}$, cf. \cite{Tsuruta:1979, ST83,Friman:1978zq}. The evaluation of the MU and the NB neutrino emissivities requires to know the nucleon-nucleon ($NN$) interaction amplitude in the baryon medium. For the  $NN$ interaction Friman and Maxwell \cite{Friman:1978zq} used the free one-pion exchange (FOPE) model. The density dependence of the reaction rates calculated with the FOPE model is very weak and thereby the neutrino radiation from a NS in this model depends very weakly on the star mass. One gets an estimate of the emissivity of the MU processes, $\epsilon_{\nu}^{\rm MU}\sim 10^{21} T_9^8$ $\frac{{\rm erg}}{{\rm s}\cdot {\rm cm}^3}$. This naive estimate used in many works is essentially modified, if various in-medium effects are taken into account in the $NN$ interaction amplitude \cite{Voskresensky:1985qg,Voskresensky:1986af,Senatorov:1987aa,Voskresensky:1987hm,
Migdal:1990vm,Voskresensky:2001fd}.

Another important point is that the nucleons may form $nn$ and $pp$ Cooper pairs, cf. \cite{Migdal1959,ST83,Senatorov:1987aa,Voskresensky:1987hm,Migdal:1990vm,Yakovlev:1999sk,
Yakovlev:2000jp,Voskresensky:2001fd,Sedrakian:2018ydt}, if the temperature is lower than corresponding critical temperatures $T^{nn}_c$ and $T^{pp}_c$, respectively.  Neutrons undergo $^1S_{0}$ pairing for $n\lsim (0.6-0.8) \, n_0$ and $^3P_2$ pairing for $0.8 \, n_0 \lsim n \lsim (3-4) \, n_0$, whereas protons are paired in $^1S_{0}$ state for $n\lsim (2-4) \, n_0$, where $n_0 = 0.16$ fm$^{-3}$ is the nucleon saturation
density. Typical values for the $nn$ and $pp$ pairing gaps, $\Delta_{nn}$ and  $\Delta_{pp}$, vary in the range  $\sim (0.1-{\rm several})$~MeV, cf. \cite{HGRR1970,Tamagaki1970} and the recent review \cite{Sedrakian:2018ydt} for further references. The critical temperature  for the $^1S_0$ pairing is  $T^{NN}_c \simeq 0.57 \Delta_{NN}$. Due to an exponential dependence on the value of the $NN$ interaction amplitude in the particle-particle channel, existing estimates of the pairing gaps are rather uncertain. The values of the $^3P_2$ neutron pairing gaps are known especially poorly. BCS-based estimates \cite{HGRR1970,Takatsuka:2004zq} yield values similar to those for the $^1S_0$ pairing. However, taking into account a medium-induced spin-orbit interaction leads to a tiny value of $\Delta({^3P_2}) \lsim 10$ KeV~\cite{Schwenk:2003bc}. Within the same scenario as in this work the dependence of the cooling curves on the values of $^3P_2$ neutron and $^1S_0$ proton paring gaps was studied in~\cite{Grigorian:2005fn}. In this work a reasonable fit of the compact star cooling data was obtained for a strongly suppressed value of the $^3P_2$ neutron pairing gap, in favour of  results of ~\cite{Schwenk:2003bc}.
The emissivities of the reactions with participation of nucleons are  suppressed by the so-called $R$-factors, describing the available phase space for a reaction. Typically the emissivity of the DU  processes is suppressed by  $e^{-(\max\{\Delta_{nn}, \Delta_{pp}\}/T)}$.  Nevertheless the DU processes remain relatively rapid even in presence of the nucleon superfluidity, because the emissivities of the MU and NB processes are suppressed even more, e.g.  for the neutron branch of the MU process a typical suppression factor is $e^{-(\Delta_{nn}+ \Delta_{pp})/T}$, cf.  \cite{ST83,Maxwell:1979zz}. In a realistic calculation the R-factors have essentially more complicated forms, cf.  \cite{Yakovlev:1999sk,Yakovlev:2000jp}, which we take into account in our scenario.

In the historically first so-called {\em{standard scenario}} of NS cooling \cite{Tsuruta:1979,ST83} the processes were calculated without taking in-medium effects into account. By that time only upper limits on NS surface temperatures were measured by the Einstein observatory at an assumption that there exist NSs in the observed supernova remnants. Some of these upper limits proved to be high, other low  but it was not known whether the NSs exist in those remnants.
For the ideal Fermi gas of nucleons \cite{Bahcall:1965zzb} the DU processes are not allowed by the energy-momentum conservation. Thereby the MU processes  were considered as the most important channel for relevant values of internal temperatures,  $T\sim 10^{8}$--$10^{9}$~K  for $t\lsim 10^5$ yr. The MU and NB processes were evaluated using the FOPE model of Friman and Maxwell \cite{Friman:1978zq}.  With the MU and NB  processes it was possible to explain the highest of measured upper limits on the surface temperatures of NSs, see Fig. 11.3 in \cite{ST83}.  To explain the lowest measured upper limits at assumption of existence of NSs in those remnants one exploited a possibility of the $P$-wave pion condensation in NS interiors suggested by A.B. Migdal, see \cite{Migdal:1978az}. In presence of a pion condensate pion Urca (PU) processes $N_1\pi_c \to N_2 e\bar{\nu}$ become possible for $n>n_c^{\rm PU}> n_0$.
Their  emissivity is roughly estimated as $\epsilon_{\nu}^{\rm PU}\sim 10^{26} T_9^6\Theta (n-n_c^{\rm PU})$ $\frac{{\rm erg}}{{\rm s}\cdot {\rm cm}^3}$ for non-superfluid systems, cf. \cite{Maxwell:1977zz}.  Like the DU processes, the PU processes are of the one-nucleon origin, but  the requirement of energy-momentum conservation can be always fulfilled for $n>n_c^{\rm PU}$ by absorbing momentum by the condensate.  In  the eighties of the previous century the prevailing opinion was that all
NS masses should be close to the values for known binary
radio pulsars, $(1.35 - 1.4)\, M_{\odot}$, cf. \cite{ST83}. Under this assumption within the standard scenario  two possibilities were considered. First one was that in supernova remnants for which  the low upper limits on the NS surface temperatures were put  there are no NSs,
and that there is no pion condensation in NSs.
Second one was that in at least some supernova remnants for which  the low upper limits on the NS surface temperatures were put there are  NSs, and thereby there is pion condensation in NSs. The existed data  did not allow to determine whether the observable objects are the NSs, slowly cooled mainly by the MU processes, or the rapid coolers with the PU processes enabled.

Besides the DU, MU and MN reactions a  nucleon pair-breaking-formation (PBF) reaction channels $N\to N\nu\bar{\nu}$, $N=n,p$, are opened up for $T<\{T^{nn}_c, T^{pp}_c\}$ respectively \cite{Flowers:1976ux,Senatorov:1987aa,
	Voskresensky:1987hm}. The nPBF neutrino process in presence of the $nn$ pairing was introduced first
in \cite{Flowers:1976ux} to occur on the vector current. Its emissivity was estimated without inclusion of in-medium effects in vertices of the process to be of the order or smaller than that for the MU processes.
References \cite{Senatorov:1987aa,
	Voskresensky:1987hm} performed calculations   of the neutrino emissivity in the non-equilibrium Green function technique that naturally allows to separate various processes within the quasiparticle approximation valid for $T\ll E_{{\rm F},N}$, where $E_{{\rm F},N}$ is the nucleon Fermi energy. The authors considered emissivity of the nPBF process  and suggested a possibility of the pPBF one in presence of the  $pp$ pairing.
They found that  the emissivity of the PBF processes is roughly estimated as $\epsilon_{\nu}^{\rm PBF}\sim 10^{28} [\frac{\Delta_{NN}}{{\rm MeV}}]^7 \sqrt{T/\Delta_{NN}} e^{-(2\Delta_{NN}/T)}$ $\frac{{\rm erg}}{{\rm s}\cdot {\rm cm}^3}$ that exceeds the emissivity of the MU processes for $\Delta_{NN}\gsim 10^9$ K. The effect of the PBF reactions on the cooling was first incorporated in the cooling code in \cite{Schaab:1996gd}. Then these processes were included in all
existing cooling codes.  Already in \cite{Senatorov:1987aa,
	Voskresensky:1987hm} an important role of in-medium effects in PBF processes was pointed out, especially for the pPBF reactions. Further, Leinson and Perez \cite{Leinson:2006gf} noticed that in the vector current channel incorporation of in-medium effects needs a special care due to a necessity to fulfill the Ward-Takahashi identities (purely in-medium effect!). Thereby
the emissivity of the  nPBF process on the vector current in presence of the $^1S_0$ $nn$ pairing proves to  be dramatically suppressed,
by a pre-factor $v_{\rm F}^4$, where $v_{\rm F}$ is the nucleon Fermi velocity.
Detailed analyses of the PBF reactions \cite{Kolomeitsev:2008mc,Kolomeitsev:2010hr,Kolomeitsev:2011wz} have shown that by taking into account the in-medium dressing of vertices, the main
contribution to the PBF emissivity in presence of the $^1S_0$ $nn$ and/or $pp$ pairings
comes from processes on the axial current, and the emissivity is thereby suppressed only as $v_{\rm F}^2$ rather than as $v_{\rm F}^4$. Also, \cite{Leinson:2011jr} noticed that in the presence of the $^3P_2$ $nn$ pairing the nPBF emissivity might be  not suppressed by the $v_{\rm F}^2$ factor,  in favor of  numerical estimates \cite{Voskresensky:1987hm} previously used in the neutron-star cooling  simulations, cf.  \cite{Schaab:1996gd}. Recall here that, if the $^3P_2$ $nn$ pairing gap is $\lsim 10$ KeV as follows from estimates ~\cite{Schwenk:2003bc}, the nPBF processes are not effective at all for $t\lsim (10^5 -10^6)$ yr of our interest.

At present time there exists information on surface temperature-time ($T_s -t$) dependence for many
pulsars. The  $T_s-t$ data can be separated in three
groups dubbed as slow cooling, intermediate cooling and rapid cooling. Recently the Chandra observatory measured the suface temperature of the young compact star in the remnant of the
historical supernova Cassiopeia A (Cas A) of the year
1680, cf. \cite{Tananbaum:1999kx,Hughes:1999ph}. The description of the cooling of this object  caused a lovely discussion \cite{Page:2010aw,Shternin:2010qi,Blaschke:2011gc,Blaschke:2013vma,
	Elshamouty:2013nfa,Ho:2014pta,Grigorian:2016leu}.
Besides, the cooling model must also explain
the hot central compact object  in the supernova remnant
XMMU J173203.3-344518, cf. \cite{Klochkov}.
Note that in order to explain the
difference in the cooling of the slowly and rapidly cooling
objects a three order of magnitude difference in their
luminosities is required \cite{Voskresensky:2001fd}.

Basing on the standard scenario the so-called  {\em{minimal cooling paradigm}} was formulated \cite{Page:2004fy,Page:2009fu}. As in the standard scenario, in this approach the medium effects are assumed to play only a minor role and masses of NSs are assumed to be close to $1.4M_{\odot}$. In this scenario the difference in the measured values of the surface temperatures of various sources of soft $X$ rays is supposed to be explained by the heterogeneity in envelope compositions for
the young stars: light element compositions for some of them and heavy element compositions for others.
A similar approach  has been extensively used by other researches, cf. \cite{Yakovlev:1999sk,Yakovlev:2000jp,Tsuruta:2002ey,Page:2010aw,Shternin:2010qi,Elshamouty:2013nfa,Ho:2014pta,Potekhin:2015qsa,Raduta:2017wpp} and refs. therein. The MU and NB  processes in these works were considered within the FOPE model of \cite{Friman:1978zq}. The nPBF processes for the $^1S_0$ pairing on the vector current were included first, as in \cite{Flowers:1976ux}, without in-medium modification factors and later the processes on the vector current were suppressed following \cite{Leinson:2006gf} and the processes on axial current were included with taking into account the  $v_{\rm F}^2$ suppression factor as in  \cite{Kolomeitsev:2008mc,Kolomeitsev:2010hr}.
The emissivity of the pPBF processes calculated without in-medium modification is very small, so these processes are considered as not effective in the minimal cooling scenario. In reality due to in-medium modification of the vertices these processes might be almost as effective as nPBF ones \cite{Kolomeitsev:2008mc,Kolomeitsev:2010hr}, depending on values of $nn$ and $pp$ pairing gaps.  The values of the $nn$ gap for the $^3P_2$ $nn$ pairing and the $^1S_0$ $pp$ gaps were found to better fit the  data on the surface temperatures of the pulsars. For the $^3P_2$ $nn$ pairing  values $\Delta_{nn}\sim 10^9$~K were suggested.
The DU processes were assumed not to occur. Thereby the main cooling agents  in this scenario are the MU, NB and nPBF processes the latter going on the  neutrons paired in the $^3P_2$ state. It proves to be that the existing data on the time dependence of the surface temperatures of pulsars are  hardly explained within the minimal cooling scenario. In particular, the hot object XMMU J173203.3-344518 is not explained. The agreement with other data can be achieved only if the scenario is supplemented by a possibility  of the efficient DU reaction   in majority of NSs. Under the outdated assumption that masses of all the NSs with measured surface temperatures are close to $1.4 \, M_{\odot}$, the DU process should be  effective already for $M\simeq 1.4 \, M_{\odot}$.  However the latter   assumption   disagrees with a broad distribution of NSs over the masses, as follows from the population synthesis modeling  and from supernova simulations \cite{Popov:2004ey,Alsing:2017bbc}.
Moreover, recent measurements of  masses of the heaviest binary pulsars demonstrated that the maximum compact star mass should be $>2 \, M_{\odot}$. It was found that PSR  J1614-2230 has the mass $M = 1.928 \pm 0.017 \, M_\odot$ \cite{Demorest:2010bx,Fonseca:2016tux} and  PSR~J0348+0432, the mass $M = 2.01 \pm 0.04 \, M_\odot$~\cite{Antoniadis:2013pzd}.
The measurements of the high masses of the pulsars  PSR  J1614-2230  \cite{Demorest:2010bx,Fonseca:2016tux}  and  PSR  J0348-0432  \cite{Antoniadis:2013pzd}   and of the low masses for PSR J0737-3039B \cite{Kramer:2006nb} for the companion of PSR
J1756-2251 \cite{Faulkner:2004ha,Ferdman} and the companion of PSR
J0453+1559   have provided the proof for the existence of
NSs  with masses varying at least from $1.2$ to $1.97 \, M_{\odot}$.
Thus, to explain existence of heaviest NSs  the EoS of the NS matter should be sufficiently stiff, cf. \cite{Klahn:2006ir,Voskresenskaya:2012np,Gandolfi:2015jma,Lattimer:2015nhk}.

Already long ago Refs. \cite{Voskresensky:1985qg,Voskresensky:1986af} suggested that different NSs have different masses and the $T_s -t$ history of NSs of various  masses should be essentially different. With this assumption  first the Einstein observatory data were explained and then it was shown  that also  Chandra observations can be naturally explained.
Refs. ~\cite{Voskresensky:1985qg,Voskresensky:1986af,Senatorov:1987aa,
	Voskresensky:1987hm,Migdal:1990vm,Voskresensky:2001fd}  suggested that for $n\gsim n_0$ the $NN$ interaction amplitude is mainly controlled by the soft pion exchange. The pion softening effect, which increases with increase of the density, is the consequence of the strong polarization of the dense nuclear matter. The authors calculated the rates of the two-nucleon  processes with inclusion of in-medium effects and found a strong dependence of the $NN$ interaction amplitude on the density (and respectively mass of the object).
Thus  the FOPE-based MU diagram
\begin{center}
{\includegraphics[width=.2\textwidth]{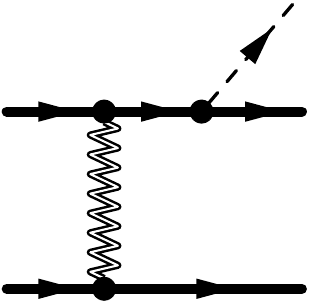}}
\label{pairneutr}
\end{center}
(the zig-zag line corresponds to the free pion and the dots denote free vertices) should be replaced by the medium one-pion exchange (MOPE) diagrams of the medium modified Urca (MMU) processes:
\begin{center}
{\includegraphics[width=.2\textwidth]{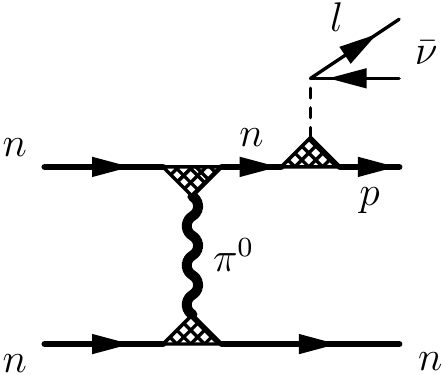}}\,,\quad \quad
{\includegraphics[width=.2\textwidth]{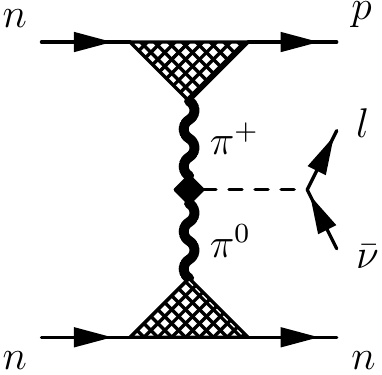}},\,\,\quad\quad
{\includegraphics[width=.2\textwidth]{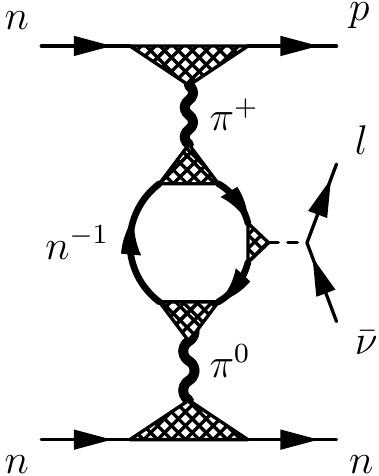}},\,\,
\label{MMU123}
\end{center}
where bold-wavy line corresponds to the dressed pion and the hatched vertex takes into account the $NN$ correlations. The first  diagram naturally generalizes the  MU (FOPE)
contribution (\ref{pairneutr}).  Besides the MOPE, the $NN$ interaction amplitude contains also a more local interaction part.
However the  correlation effects on the local $NN$ interaction lead to a suppression of the amplitude. Therefore for $n\gsim n_0$ the main contribution is given by the MOPE. As the result of the pion softening, the $NN$ interaction amplitude proves to be strongly enhanced for $n>n_0$, \cite{Voskresensky:1986af,Migdal:1990vm,Voskresensky:2001fd}, whereas
for $n\ll n_0$ the same polarization  effects result in a suppression of the $NN$ amplitude compared to that computed in the FOPE model, cf. \cite{Blaschke:1995va,Hanhart:2000ae,Knoll:1995nz}. Evaluations  have shown that first diagram (\ref{MMU123}) gives a smaller contribution to the MMU emissivity for $n \gsim n_0$ than the second and third diagrams, which incorporate  processes occurring through intermediate reaction states.   Note that the latter two diagrams do not contribute, if one approximates the nucleon-nucleon interaction  by a two-body potential. Also  the second and third  diagrams (\ref{MMU123}) do not contribute to the emissivity of the medium-modified nucleon bremstrahlung (MNB) processes due to symmetry arguments. All these peculiarities of the density dependence of the $NN$ interaction amplitude were taken into account in the mentioned works, see \cite{Voskresensky:1986af,Migdal:1990vm,Voskresensky:2001fd}.

For $n=n_c^{\rm PU}$, the pion condensation may appear by a first-order phase transition, see \cite{Migdal:1990vm}. Estimates give $n_0<n_c^{\rm PU}\lsim 3n_0$.
Evaluations ~\cite{Voskresensky:1986af,Senatorov:1987aa,
	Voskresensky:1987hm,Migdal:1990vm,Voskresensky:2001fd} showed that at $n=n_c^{\rm PU}$ the MMU emissivity may exceed the MU one calculated with the FOPE model by $3-4$~orders of magnitude.
Besides the MMU and MNB processes \cite{Voskresensky:1985qg,Voskresensky:1986af}, the PBF processes \cite{Senatorov:1987aa,
	Voskresensky:1987hm,Kolomeitsev:2008mc,Kolomeitsev:2010hr,Kolomeitsev:2011wz} and other ones were incorporated. Basing on results of these works  the {\em{nuclear medium cooling scenario}} was developed. The results of the calculations were confronted to the  data on the $T_s-t$ plane \cite{Schaab:1996gd,Blaschke:2004vq,Grigorian:2005fn,Blaschke:2011gc,Blaschke:2013vma,
	Grigorian:2016leu} demonstrating an overall agreement, without necessity to assume presence of the DU processes. Moreover, the absence of sub-millisecond pulsars might be  naturally explained within the nuclear medium cooling scenario \cite{Kolomeitsev:2014epa,Kolomeitsev:2014gfa}. Two possibilities, one which allows for the PU processes and other one not allowing the pion condensation, were considered. In both cases the $T_s-t$ data can be appropriately described. Near the pion condensation threshold there may also exist a region of the fermion condensation that allows for the neutrino processes with the neutrino emissivity $\propto T^5$, cf. \cite{Voskresensky:2000px}. Besides the $P$-wave pion condensation, in the dense matter may occur the $S$-wave or $P$-wave kaon condensations \cite{Kaplan:1987sc,Kolomeitsev:1995xz,Kolomeitsev:2002pg} and the $S$-wave charged $\rho$ condensation \cite{Voskresensky:1997ub,Kolomeitsev:2004ff,Kolomeitsev:2017gli}. Although all these possibilities lead to an enhanced cooling of NSs \cite{Tatsumi:1988up,Fujii:1993cf,Kolomeitsev:2004ff} the resulting emissivities are typically hundred times smaller than that for the DU process. For the sake of simplicity we focus below on the possibility that the pion softening is saturated for $n\gsim n_c^{\rm PU}\sim 3n_0$ and disregard the possibilities of pion, kaon, charged $\rho$-meson and fermion condensations.

First simulations done within  the nuclear medium cooling scenario \cite{Schaab:1996gd,Blaschke:2004vq,Grigorian:2005fn,Blaschke:2011gc} exploited the HHJ version of the variational Akmal-Pandharipande-Ravenhall (APR) EoS with a fitting parameter $\delta = 0.2$ \cite{Heiselberg:1999mq} yielding $M_{max}\simeq 1.92 \, M_{\odot}$. In \cite{Blaschke:2013vma} we  exploited the HDD EoS, being similar to the APR EoS \cite{Akmal:1998cf} up to  $4\,n_0$, but  stiffer at higher densities than the HHJ EoS, producing the maximum NS mass compatible with the observations of PSR  J1614-2230 \cite{Demorest:2010bx,Fonseca:2016tux}  and  PSR  J0348-0432  \cite{Antoniadis:2013pzd}.  In \cite{Grigorian:2016leu} we used still stiffer DD2 and DD2vex EoSs
satisfying the constraint $M_{max}>2 \, M_{\odot}$, cf. \cite{Typel:2009sy}. The DD2 EoS and its modification DD2vex EoS are the RMF based EoSs  where one uses the  density dependence of the  couplings. However  all these EoSs do not include the possibility of filling of the hyperon Fermi seas.

Appearance of hyperons leads to a softening of
the EoS and reduction of the maximum NS mass. By employing a
hyperon-nucleon potential, the maximum mass
of NSs with hyperons was computed  ~\cite{Djapo:2008au} to be well below $1.4
M_{\odot}$. Within the RMF approach the critical densities for the appearance of first hyperons prove to be rather low, $n_{\rm c}^{H}\sim 3n_0$, cf.~\cite{Glendenning,SchaffnerBielich:2008kb,Weissenborn:2011kb}, if the hyperon coupling constants satisfying the SU(6) symmetry relations  are fitted from the hyperon potentials in the nuclear medium at $n= n_0$. The
difference between NS masses  obtained with and without inclusion of hyperons proves to
be so large for reasonable hyperon fractions in the standard RMF
approach, that in order to get the maximum NS mass satisfying  experimental constraints one has to start with
very stiff purely nucleon EoS.  The latter assumption hardly agrees with the results
of the microscopically-based variational EoS~\cite{Akmal:1998cf} and the
EoS calculated with the help of the auxiliary field diffusion
Monte Carlo method~\cite{Gandolfi:2009nq}. The problem  was called ``the hyperon puzzle"  \cite{SchaffnerBielich:2008kb}. The suggested
explanations required additional assumptions, see discussion
in~\cite{Fortin:2014mya}, for example, the inclusion of an
interaction with a $\phi$-meson mean field, and the usage of
smaller hyperon-nucleon coupling constants following the SU(3)
symmetry relations~\cite{Weissenborn:2011ut}, as well as other
modifications.
The above mentioned hyperon puzzle is naturally solved
within  the RMF  EoSs with hadron effective masses and coupling constants dependent on the scalar field $\sigma $ \cite{Maslov:2015msa,Maslov:2015wba}. Appropriate KVORcutH$\phi$ and  MKVORH$\phi$ models were constructed. Other important constraints on the EoS, e.g. the flow constraint from heavy-ion collisions \cite{Danielewicz:2002pu,Fuchs,Lynch:2009vc}, are fulfilled within these models as well \cite{Maslov:2015msa,Maslov:2015wba,Kolomeitsev:2016ptu,Kolomeitsev:2017gli}.

However, there remains another part of the hyperon puzzle: in presence of  hyperons  the  efficient DU reactions on hyperons, e.g. $\Lambda\to p+e+\bar{\nu}$ become possible, cf. \cite{Maxwell:1986pj}. These reactions are operative only, if the hyperon concentration exceeds some threshold value, that may happen with an increase of the baryon density $n$ above some critical value $n^{\rm DU}_{c, H}$. They accelerate the cooling of NSs with $M>M^{\rm DU}_{c, H} $, where $M^{\rm DU}_{c, H} $ is a NS mass, at which the central density reaches the critical value $n^{\rm DU}_{c, H}$.
In the reaction on $\Lambda$ mentioned above the threshold value is only slightly above the density $n_{c,\Lambda}$ of their appearance.
Up to now the cooling of NSs with inclusion of hyperons with a stiff EoS satisfying the constraint $M_{max}>2 \, M_{\odot}$  was considered  only within the minimal cooling scenario, cf. \cite{Raduta:2017wpp}. As it is seen from the figures of this work,  the existing surface
temperature-age data are hardly described in the minimal cooling scheme.

In the given paper we demonstrate how a satisfactory explanation of  existing observational pulsar cooling data is reached within  the ``nuclear medium cooling"
scenario of \cite{Blaschke:2004vq}, now  with  the RMF EoS MKVORH$\phi$ with a $\sigma$-scaled hadron effective masses and coupling constants including hyperons \cite{Maslov:2015msa,Maslov:2015wba}. The paper is organized as follows. In the next section we introduce the MKVORH$\phi$  EoS which we then employ in calculation of the cooling history of NSs. Section \ref{sect::cool_input} introduces inputs for the neutrino cooling calculations. Section \ref{sect::results} presents results of numerical calculations. Section \ref{sect::conclusion} contains concluding remarks. The results were briefly announced on the conference ``Compact Stars in the QCD Phase Diagram VI: Cosmic matter in heavy-ion collision laboratories?" (CSQCD VI), cf. \cite{Grigorian:2017xqd}.

\section{Equation of state}
\label{sect::eos}
An EoS of cold hadronic matter should
satisfy empirical constraints on global characteristics of atomic nuclei;
constraints on the pressure of the nuclear mater  from the description of particle transverse and elliptic flows  in heavy-ion collisions, cf.  \cite{Danielewicz:2002pu,Lynch:2009vc};
allow for the heaviest known   pulsars PSR  J1614-2230  and  PSR~J0348+0432~\cite{Fonseca:2016tux,Antoniadis:2013pzd};
allow for an adequate description of the  cooling of NSs, most probably without DU neutrino processes in the majority of the known pulsars detected in soft $X$ rays, cf.  \cite{Klahn:2006ir};
yield a mass-radius relation compatible with empirical constraints including the recent gravitation wave LIGO-Virgo detection  GW170817  \cite{TheLIGOScientific:2017qsa, Annala:2017llu};
 pass the constraint on the  relation between tidal deformabilities of two merging objects  following from the analysis of GW170817 \cite{Abbott:2018exr};
being extended to non-zero temperature $T$ (for $T<T_c^{\rm QH}$ where $T_c^{\rm QH}$ is the critical temperature of the  possible quark-hadron phase transition)  appropriately describe supernovae and  matter of proto-NSs,  heavy-ion collision data, and lattice data, cf. \cite{Khvorostukhin:2006ih,Khvorostukhin:2008xn}, etc.
The most difficult  is to satisfy simultaneously the heavy-ion-collision flow and the maximum NS mass constraints, since the fulfillment of the flow constraint~\cite{Danielewicz:2002pu,Lynch:2009vc} requires a rather soft EoS of the isospin-symmetric matter (ISM), whereas the EoS of the beta-equilibrium  matter (BEM) should be stiff in order to predict the maximum mass of a NS  higher than the measured mass $M = 2.01 \pm 0.04 \, M_\odot$~\cite{Antoniadis:2013pzd}  of the  pulsar PSR~J0348+0432, being the heaviest among the known pulsars.

Reference  \cite{Kolomeitsev:2004ff} suggested a set of RMF models with the hadron masses and meson-baryon coupling constants  dependent on the scalar mean field $\sigma$. A working model MW($n.u., z=0.65$) labeled in \cite{Klahn:2006ir}   as KVOR model   satisfied appropriately the majority of experimental constraints known to that time.  However hyperons were not included in the EoS for $T= 0$. Without hyperons the KVOR EoS supplemented with the BPS crust EoS \cite{Baym:1971pw} yields the maximum NS mass of $2.01\,M_{\odot}$ that fits the constraints \cite{Fonseca:2016tux,Antoniadis:2013pzd}, but only marginally.
References \cite{Maslov:2015msa,Maslov:2015wba,Maslov:2015lma}  proposed modifications of the KVOR model, which allow for a better fulfillment of the existing experimental constraints. One extension of the model (KVORcut) demonstrates that the EoS stiffens, if the  growth of the scalar mean field  with an increase of the baryon density is limited after some value for density exceeding a certain quantity above $n_0$, cf. \cite{Maslov:2015lma}. This can be realized, e.g., if the nucleon - vector-meson coupling constant changes rapidly as a function of the scalar field  above the desired value. The other version of the model (MKVOR) assumes a smaller value of the nucleon effective mass at the nuclear saturation density and uses a saturation of the scalar field in the NS matter induced by a strong variation of the nucleon - isovector-meson coupling constant as a function of the scalar field.   Hyperons ($H$) were included in the KVORcutH$\phi$ and MKVORH$\phi$ extensions of the models. At the price of the choosing of an appropriate $\phi$ meson scaling function, the resulting EoSs fulfill a majority of known empirical constraints including that the maximum mass of the NSs should enlarge $2M_{\odot}$.  However, the cooling of NSs with these models was not yet studied.

\begin{figure}\centering
	\includegraphics[width = \textwidth]{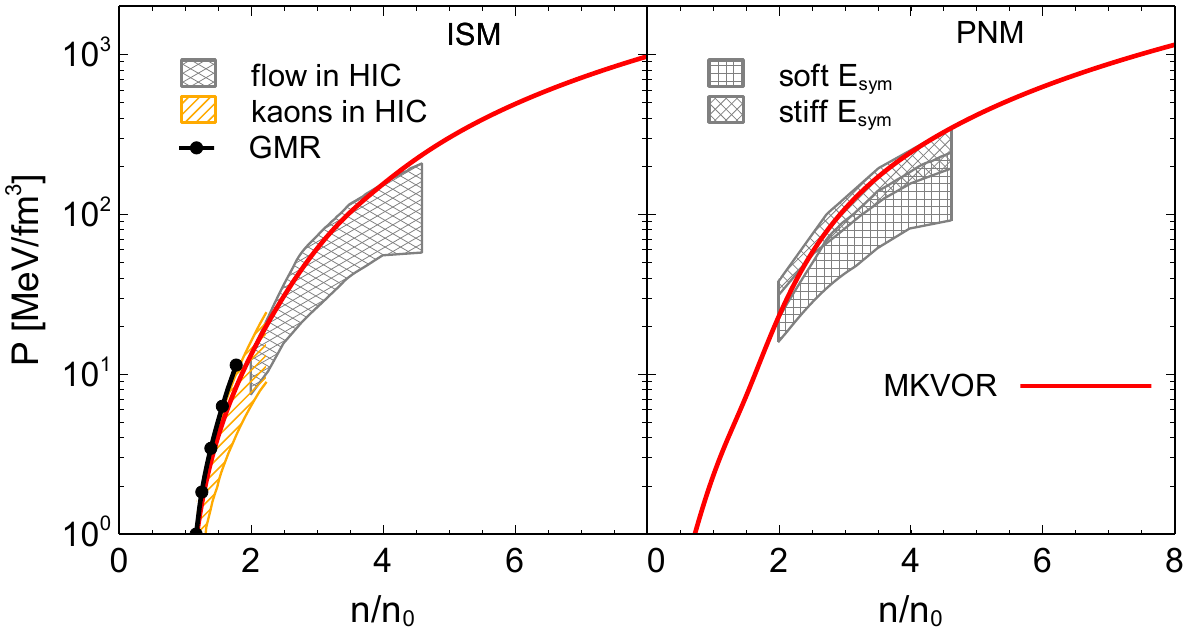}
	\caption{  Pressure  as a function of the nucleon density  for MKVOR model (coincides with that for MKVORH$\phi$ model) for ISM  (left panel) and for the PNM (right panel). Double-hatched area on left panel is the constraint from the particle flow in heavy-ion collisions \cite{Danielewicz:2002pu}, hatched area is the kaon flow constraint extracted in \cite{Lynch:2009vc} from the analysis of \cite{Fuchs}. Line with bold dots shows the extrapolation of the pressure consistent with the description of the giant monopole resonances (GMR), cf. \cite{Lynch:2009vc}. On the right panel two double-hatched areas show the pressure consistent with the flow data after inclusion of the isospin asymmetry terms with stiff and soft density dependencies. }
	\label{PressureDaniel}
\end{figure}

In the given paper we will use the EoS of the MKVOR and MKVORH$\phi$ models. In Fig. \ref{PressureDaniel} we show the pressure of the MKVOR model  as a function of the nucleon density for ISM on the left panel  and for purely neutron matter (PNM) on the right panel, see Fig. 16 in \cite{Maslov:2015wba}. In cases of ISM and PNM the MKVOR and MKVORH$\phi$ models coincide. Double-hatched area on the left panel is the constraint from the particle flow in heavy-ion collisions \cite{Danielewicz:2002pu}, hatched area is the kaon flow constraint extracted in \cite{Lynch:2009vc} from the analysis of \cite{Fuchs}. Line with bold dots shows the extrapolation of the pressure consistent with the description of the giant monopole resonances (GMR), cf. \cite{Lynch:2009vc}. On the right panel two double-hatched areas show the pressure consistent with the flow data after inclusion of the isospin asymmetry terms with stiff and soft density dependencies.

\begin{figure}
	\centering
	\includegraphics[width=.47\textwidth]{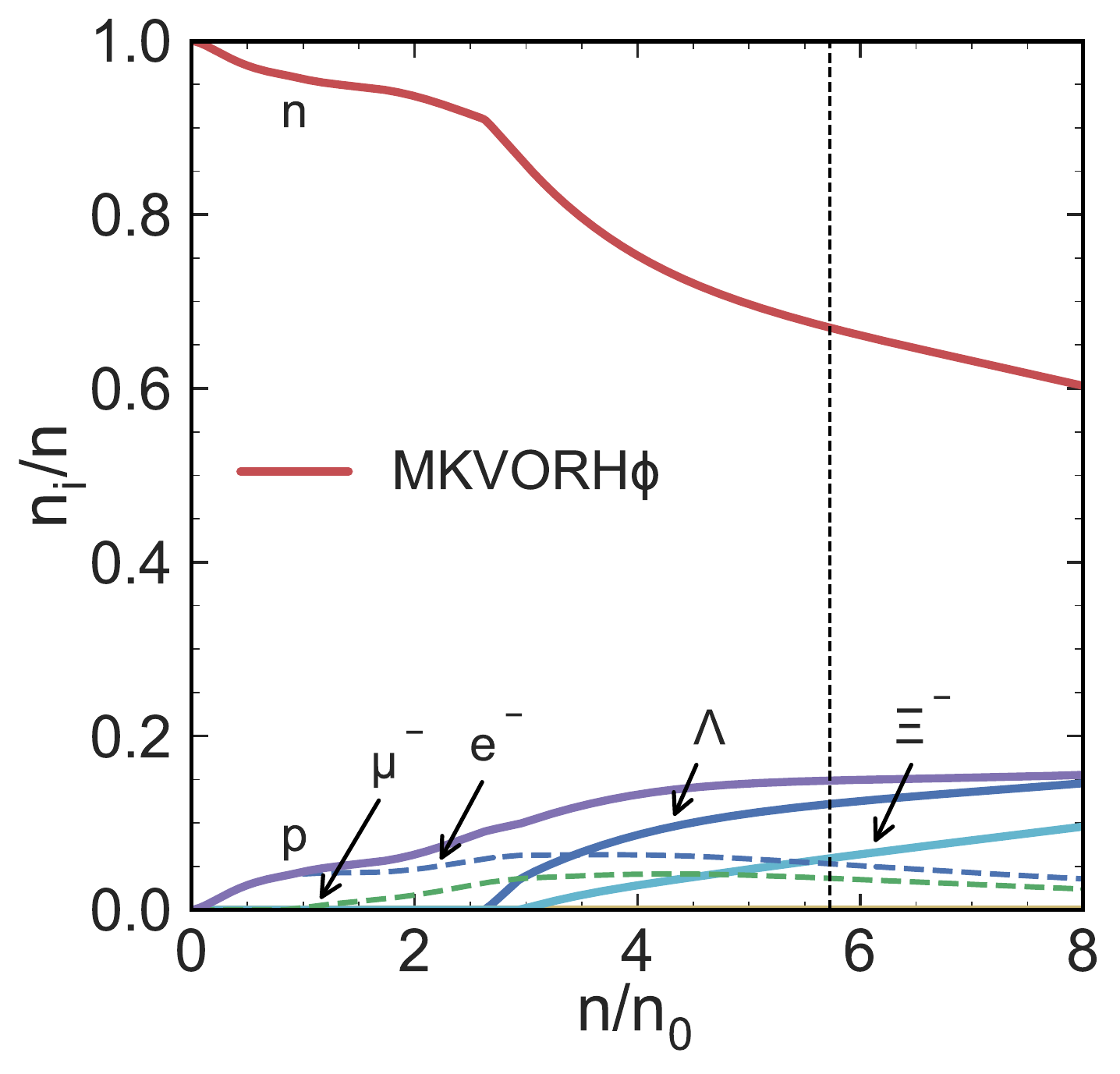}
	\includegraphics[width=.48\textwidth]{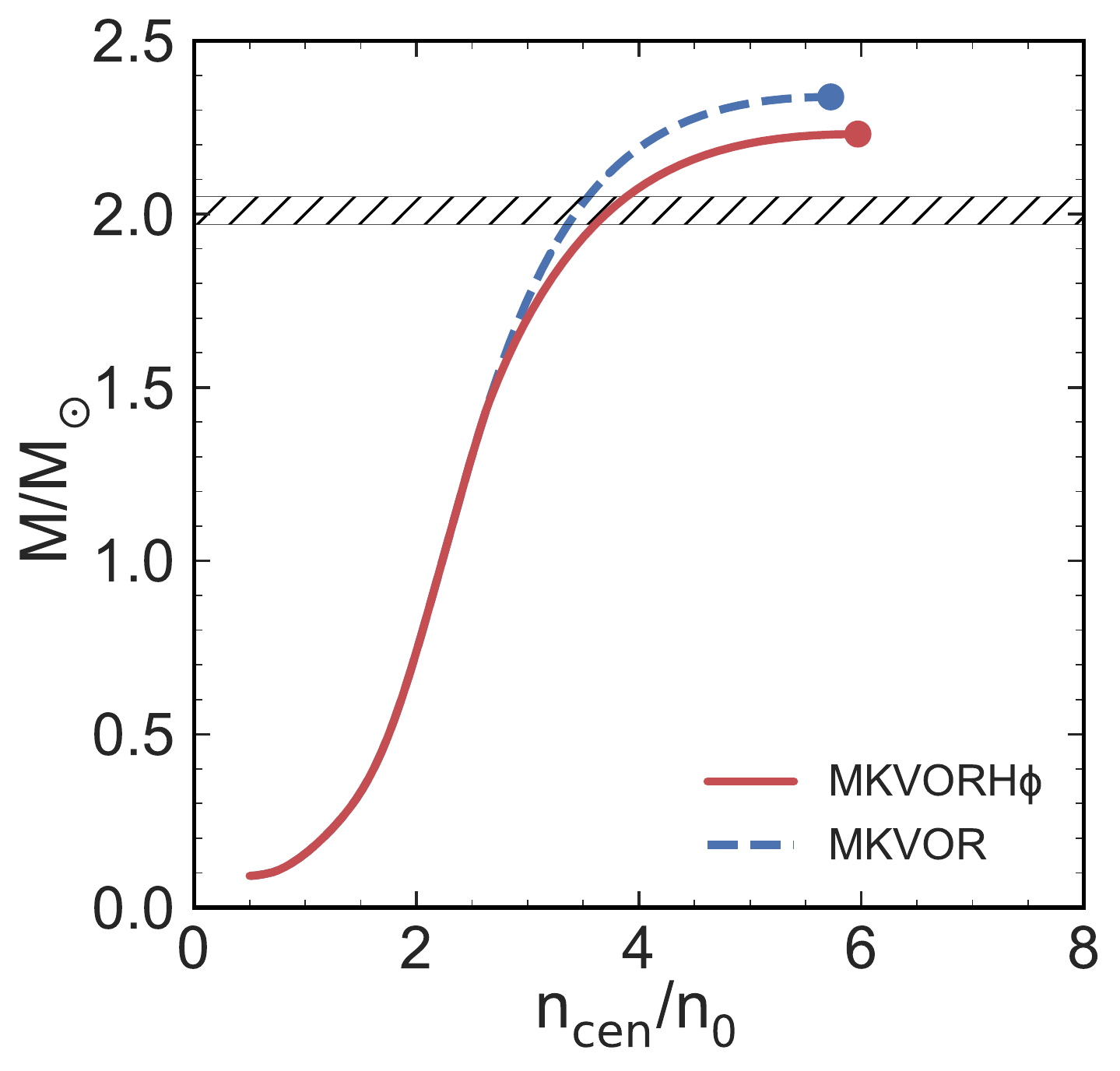}
	\caption{ Left panel: Particle concentrations as functions of the total baryon density in BEM for model MKVORH${\phi}$  (solid lines for baryons and dashed lines for leptons).
Vertical  dashed line indicates maximally possible value of the NS central density. Right panel: NS masses versus the central total baryon density for the MKVOR model without inclusion of hyperons (dashed line) and for the MKVORH$\phi$ model with included hyperons (solid line). Hatched band indicates  the measured mass $M = 2.01 \pm 0.04 \, M_\odot$~\cite{Antoniadis:2013pzd}  of the  pulsar PSR~J0348+0432.
}
\label{mkv_mass}
\end{figure}

In Fig.~\ref{mkv_mass} on left panel we show the baryon concentrations as functions of the baryon density in BEM for the MKVORH$\phi$ model being smoothly matched with the BPS EoS for the crust, see Fig. 25 (left) in \cite{Maslov:2015wba}. If we continued to use the  MKVOR model up to very low densities it almost would not affect the NS mass  but would affect the radius. With an increase of the density the  $\Lambda$ hyperons appear first at $n = 2.621 \, n_0$, and then for $n = 2.929 \, n_0$ the $\Xi^-$ hyperons arise. The $\Xi^0$ and $\Sigma$ hyperons do not appear for $n < 8 \, n_0$. The central density exceeds $2.621 \, n_0$ for
$M > 1.426 \, M_{\odot}$. The DU reactions on $\Lambda$ hyperons, $\Lambda \to p+e+\bar{\nu}$, $p+e\to \Lambda +\nu,$  become allowed for densities exceeding $2.625 \, n_0$. The central density exceeds $2.625 \, n_0$ for
$M>M_{c,\Lambda}^{\rm DU} \simeq 1.429 \, M_{\odot}$.

Thus the ``strong DU constraint" for the DU reactions on  $\Lambda$ hyperons is not fulfilled in the MKVORH$\phi$ model under consideration.  However we should bear in mind that the neutrino emissivity in the DU processes on hyperons is typically smaller than that in the standard DU processes on nucleons due to a smaller coupling constant for the hyperons (0.0394 factor  for the DU process $\Lambda \to p+e+\bar{\nu}$ and 0.0175 for $\Xi^-\to \Lambda +e+\bar{\nu}$ compared to 1 for the DU process on nucleons). For the MKVORH$\phi$ model the DU reaction on nucleons proves to be allowed for $M>M_c^{\rm DU}\simeq 2.078 \, M_{\odot}$, whereas for MKVOR model it was allowed for $M>M_c^{\rm DU}\simeq 2.149 \, M_{\odot}$. The DU reactions with participation of $\Xi^-$, $\Xi^-\to \Lambda +e+\bar{\nu}$ and $\Lambda +e\to \Xi^{-} +\bar{\nu}$, become allowed for the MKVORH$\phi$ model only for $M>M_{c,\Xi^{-}\Lambda}^{\rm DU} \simeq 1.664 \, M_{\odot}$, when $n_{\rm cen}$ exceeds the value $2.943 \, n_0$ and $\Xi^- \to n+ e + \bar \nu$ is not operative for all the densities under consideration. Also, we should bear in mind that in superfluid matter the pairing suppression $R$-factors for the DU processes on nucleons and hyperons are different.  In Fig~\ref{mkv_mass} on the right panel we demonstrate the NS mass as a function of a central density for the   MKVORH$\phi$ model, which includes hyperons (solid line) and for MKVOR model without hyperons (dashed line), cf. Fig. 25 (right) in~\cite{Maslov:2015wba}. For the MKVOR model the maximum NS mass reaches $2.338 \, M_{\odot}$. For MKVORH$\phi$ model including hyperons the maximum NS mass is only slightly smaller, $2.231 \, M_{\odot}$.

\begin{figure}
	\centering
	\includegraphics[width=.48\textwidth]{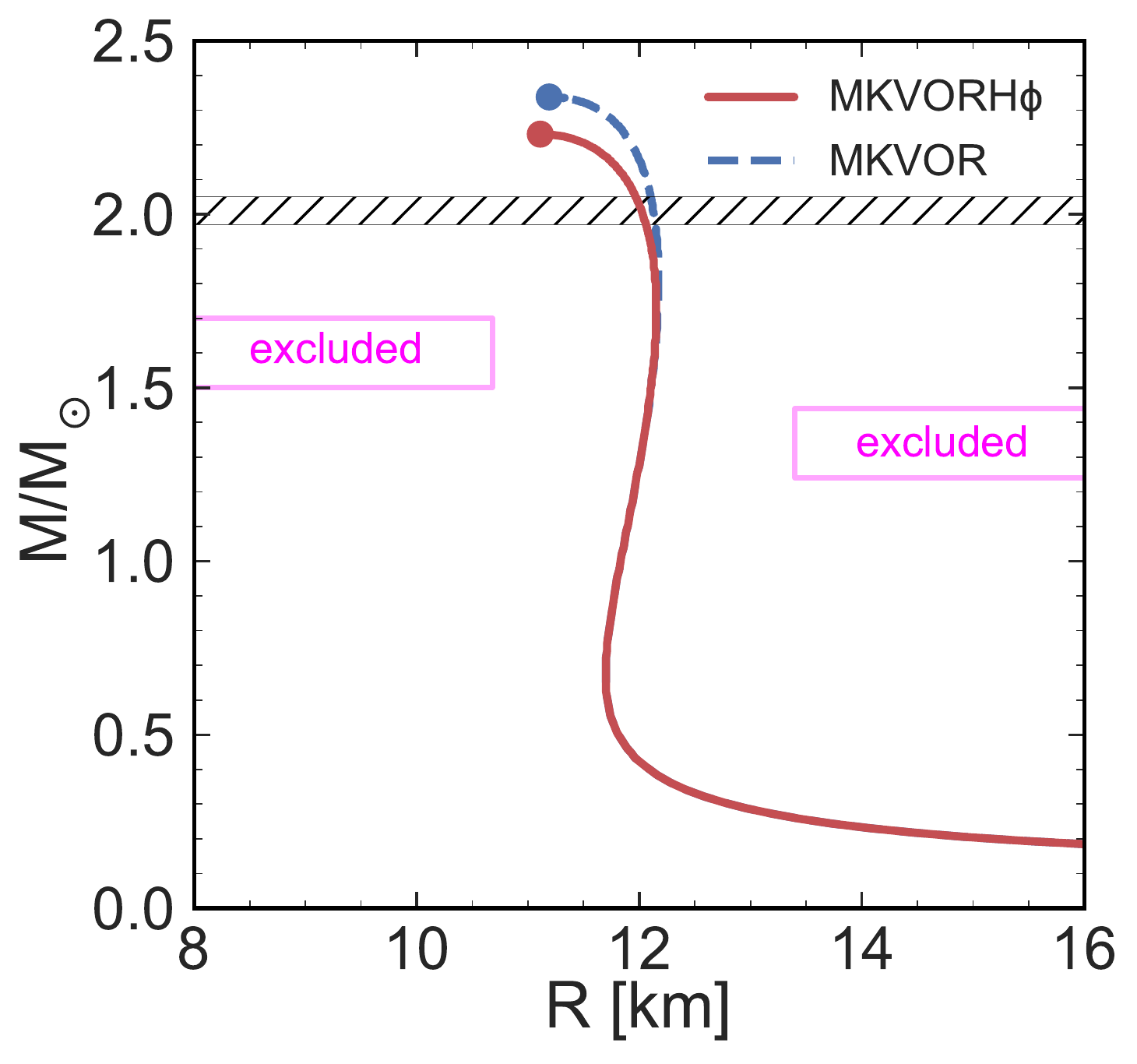}
	\includegraphics[width=.48\textwidth]{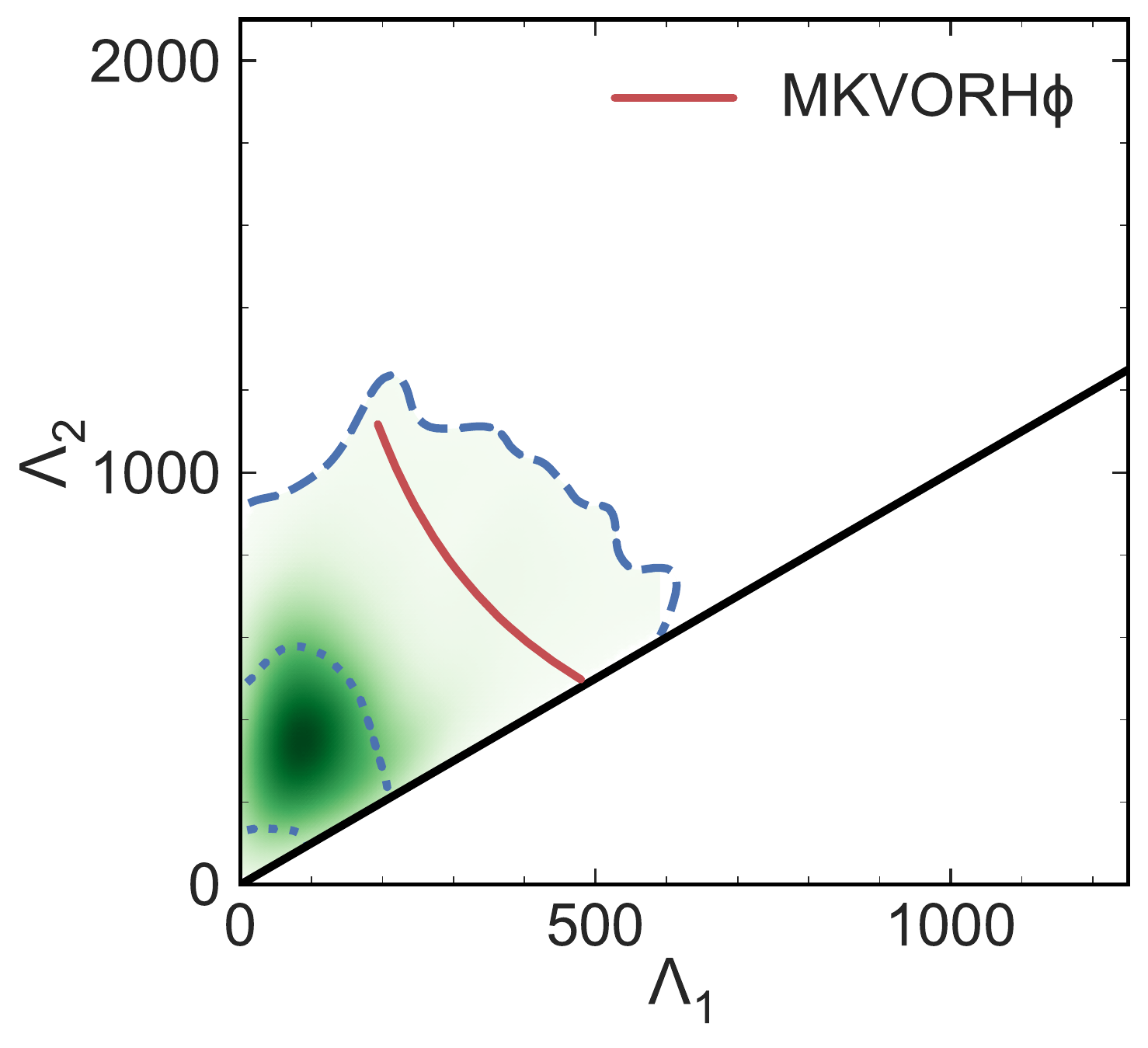}
	\caption{ Left panel: NS mass-radius relation for our MKVORH$\phi$ and MKVOR models (matched with the BPS model for the crust). Hatched band indicates  the measured mass $M = 2.01 \pm 0.04 \, M_\odot$~\cite{Antoniadis:2013pzd}  of the  pulsar PSR~J0348+0432. Two recently excluded regions $R < 10.68$ km for $1.6 \, M_\odot$ NS \cite{Bauswein:2017vtn} and $R > 13.4$ km at the NS mass of $1.4 \, M_\odot$ \cite{Annala:2017llu} are shown by  rectangles.  Right panel: Relation between tidal deformabilities $\Lambda_1$ and $\Lambda_2$ of two participants of GW170817 event for our model (solid curve), together with the probability distribution for $(\Lambda_1, \Lambda_2)$ obtained in \cite{Abbott:2018exr}  using the EoS-insensitive relation under assumption of a common EoS for two objects (shaded area). Straight line $\Lambda_1 =\Lambda_2$. Dashed and dotted lines denote the 90\% and 50\% credible levels for this EoS-insensitive distribution, respectively.}
	\label{mkv_ligo}
\end{figure}


In Fig.~\ref{mkv_ligo} we demonstrate how much NS properties in the MKVOR and MKVORH$\phi$ models matched with BPS EoS in crust are consistent with new constraints following from the GW170817 event. On the left panel we compare the predicted NS mass-radius relation with the constraints for the NS radii $R < 10.68$ km for $M = 1.6 \, M_\odot$ \cite{Bauswein:2017vtn} and $R > 13.4$ km for $M = 1.4 \, M_\odot$ \cite{Annala:2017llu} shown by the rectangles. The predictions of both our models MKVOR and MKVORH$\phi$ are consistent with these constraints. Horizontal dashed band demonstrates the same maximum NS mass constraint as in the Fig.~\ref{mkv_mass}.
 On the right panel the constraint on the tidal deformability is formulated as a region in the $\Lambda_1 - \Lambda_2$ plane. The measured chirp mass of a signal is ${\cal M} = (M_1 M_2)^{3/5} (M_1 + M_2)^{-1/5} =  \, 1.188 \, M_\odot$, and the possible ranges of participant masses are $M_1 \in (1.36-1.6)\, M_\odot$ and $M_2 \in (1.17 - 1.36) \, M_\odot$. The shaded area denotes the probability distribution for $(\Lambda_1,\Lambda_2)$ obtained in \cite{Abbott:2018exr} within an assumption of a common EoS for the two merging objects. The dashed and dotted lines denote the borders of 90\% and 50\% confidence regions, as obtained in \cite{Abbott:2018exr}, respectively. The results for both of our EoSs MKVOR and MKVORH$\phi$ (see solid line in Figure) are visually indistinguishable and fit within the 90\% confidence region. Thus the model we employ is compatible with the new constraints following from the GW170817 signal.

\section{Inputs for the cooling model}
\label{sect::cool_input}
We adopt all cooling inputs such as the neutrino
emissivities, specific heat, crust properties, etc., from our previous works   performed first with the HHJ EoS \cite{Blaschke:2004vq,Grigorian:2005fn,Blaschke:2011gc}, then with a stiffer  HDD EoS \cite{Blaschke:2013vma}, and with even more stiffer DD2 and DD2vex  EoSs \cite{Grigorian:2016leu} for the hadronic matter. These works exploit the nuclear medium cooling scenario formulated in \cite{Blaschke:2004vq},  where the most efficient  are the MMU
processes, $nn\to npe\bar{\nu}$ and $np\to ppe\bar{\nu}$, the MNB processes, $nn\to nn\nu\bar{\nu}$, $np\to np\nu\bar{\nu}$, $pp\to pp\nu\bar{\nu}$, and the nPBF and pPBF processes, $n\to n\nu\bar{\nu}$
and $p\to p\nu\bar{\nu}$, the latter going only in superfluid matter.

In
\cite{Blaschke:2004vq,Grigorian:2005fn,Blaschke:2011gc,Blaschke:2013vma,Grigorian:2016leu}
we have demonstrated that the cooling history is  sensitive to the
efficiency of the MMU processes controlled by the
density dependence of the square of the effective pion gap $\omega^{*2} (k_m)$, where
\begin{eqnarray}
\omega^{*2} (k)=-D^{R-1}_{\pi}(\mu_\pi ,k, n)=k^2 +m_\pi^2 -\mu_\pi^2 +\mbox{Re}\Pi_\pi (\mu_\pi ,k, n)\,,
\end{eqnarray}
$D^R_{\pi} (\mu_\pi ,k, n)$ is the retarded pion Green function  and $\Pi_\pi (\mu_\pi ,k, n)$ is the pion polarization operator in the BEM.
For the densities larger than some value $n_{c1}<n_0$   the quantity $\omega^{*2} (k)$  gets a minimum for the momentum  $k=k_m (n)\simeq p_{{\rm F},n}(n)$, where $p_{{\rm F},n}$ is the neutron Fermi momentum. Evaluations \cite{Voskresensky:1993ud} performed for ISM produce $n_{c1}\simeq (0.5-0.8) \, n_0$.  For the BEM under consideration we use $n_{c1}\simeq 0.8 \, n_0$, cf. \cite{Migdal:1990vm}.
For $\pi^0$ the pion frequency  $\mu_\pi =0$, for $\pi^-$ one has $\mu_\pi =\mu_n -\mu_p>0 $, where $\mu_n$ and $\mu_p$ are the neutron and proton chemical potentials. For the isospin-asymmetric matter the values $\omega^{*2} (k_m)$ for $\pi^-$ and $\pi^0$ are different.  The value $\omega^{*2} (k_m)$
enters the $NN$ interaction amplitude and the emissivity of the MMU  and MNB processes instead of the quantity $m_\pi^2 + p_{\rm F}^2$, which determines  the $NN$ interaction amplitude and the emissivity of the MU and NB processes in the minimal cooling scheme. The ratio $\omega^{*2} (k_m(n))/[m_\pi^2 + p_{\rm F}^2(n)]<1$ for $n>n_{c1}$  is a measure of the effect of the pion softening.

\begin{figure}
	\centering
	\includegraphics[height=5.5cm,clip=true]{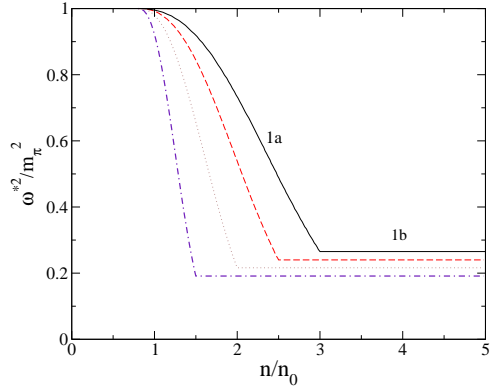}
	\caption{Density dependence of the squared effective pion gap  used in the given work. We assume that the pion softening effect saturates above a critical density, which value, $n_c^{\rm PU}$, we vary.  Broken  dash-dotted line corresponds to  $n_c^{\rm PU}=1.5 \, n_0$, dotted line corresponds to $n_c^{\rm PU}=2 \, n_0$, dashed line, to $2.5 \, n_0$, and solid line, to  $3 \, n_0$. The 1b continuation of 1a line demonstrates saturation of the pion softening for $n>n_c^{\rm PU}$.}
	\label{piongap}
\end{figure}

The effective pion gap that we use in the given work is shown in Fig. ~\ref{piongap}. With an increase of the density $\omega^{*2} (k_m(n))$ decreases for $n>n_{c1}$  reaching its positive minimum at the critical point of the first-order phase transition to the pion-condensed state.  The critical densities of the $\pi^-$ and $\pi^0$ condensate appearance $n_c^{\pi^-}$ and $n_c^{\pi^0}$, respectively, are model dependent. Simplifying we assume $n_c^{\pi^-}=n_c^{\pi^0}=n_c^{\rm PU}$ and exploit the same $\omega^{*2} (k_m(n))$ for $\pi^-$ and $\pi^0$, as we have done in our previous works, see  Fig. 2 in \cite{Grigorian:2016leu}, and we vary the value $n_c^{\rm PU}$ in  interval $(1.5 - 3) \, n_0$.   Note that following the variational calculations of Ref.~\cite{Akmal:1998cf} the neutral pion condensation in the NS matter with the APR EoS appears already for $n>n_c^{\rm PU}\simeq 1.3 \, n_0$, in favor of an even steeper  dependence $\omega^*(n)$ than that we use. Just in order to be as conservative as possible we continue to exploit a  weaker pion softening in our calculations.
Note that pion condensation can appear by the first-order phase transition only if one takes into account pion fluctuation diagrams, cf. \cite{Dyugaev:1975dk,Migdal:1990vm}.  In the critical point the quantity $\omega^{*2} (k_m (n))$ jumps then to a negative value,  $\omega^{*2} (k_m (n))<0$, describing the squared amplitude of the condensate field.
For simplicity in the given work, as in \cite{Grigorian:2016leu}, we assume a saturation of the pion softening effect  for $n>n_c^{\rm PU}$, disregarding thereby a possibility of the pion condensation (see horizontal lines  in Fig. ~\ref{piongap}).

The cooling curves $T_s (t)$ prove to be rather insensitive to the value of the $^1S_0$  $nn$ pairing gap since the $^1S_0$ neutron pairing does not spread in the interior region of  NSs with $M> 1\,M_{\odot}$. Thereby we use here the same values for $^1S_0$  $nn$ pairing gaps as  in our previous works, e.g. see Fig. 5 in \cite{Blaschke:2004vq} for details. Also, within our scenario we continue to exploit assumption that   the value of the $^3P_2$ $nn$ pairing gap is  tiny ($<10$ KeV), and thereby its actual value does not affect the calculations of the neutrino emissivity \cite{Grigorian:2005fn}.

\begin{figure}
	\centering
	\includegraphics[width=0.45\textwidth]{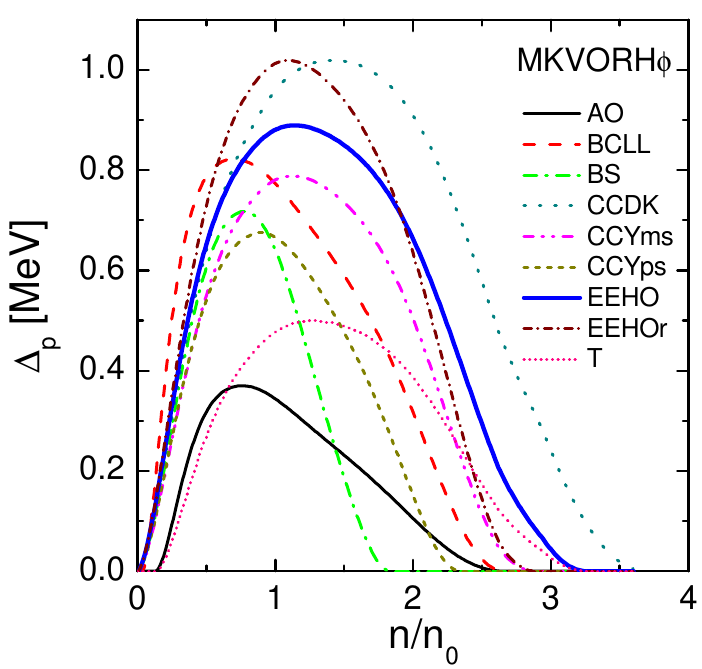}
	\includegraphics[width=0.45\textwidth]{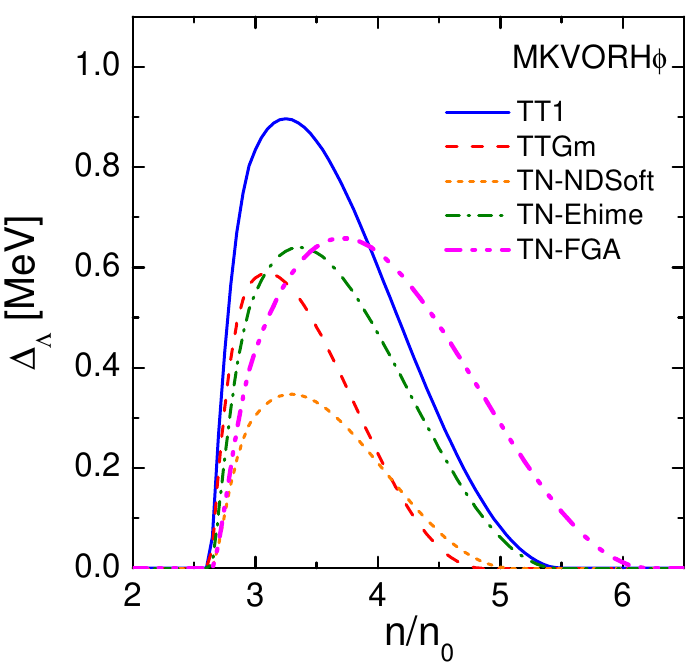}
	\caption{Pairing gaps for protons (left panel) and $\Lambda$ hyperons (right panel) for $T=0$ as functions of the baryon density for the  MKVORH$\phi$ EoS including hyperons. Proton gaps are evaluated using the same models as in~\cite{Ho:2014pta,Grigorian:2016leu} and the $\Lambda$ hyperon gaps are from~\cite{TT00,TN06,LanskYam}.}
	\label{Protongaps}
\end{figure}

The results are sensitive to the values of the proton $^1S_0$ pairing gaps since these gaps, being calculated in many works with various EoSs, are not  small (typical values are  $\sim (0.1-1)\,{\rm MeV}$) and they spread into the interior region up to densities  $n\sim (2-4)\,n_0$.  We use the parametrization of the zero-temperature $pp$ pairing gaps
$\Delta_p (p_{{\rm F},p})$ from \cite{Ho:2014pta}, Eq. (2);  $p_{{\rm F},i}$ denotes the
Fermi momentum of the species $i$.
The parameters are taken to fit the gaps computed in various publications.
The abbreviations of the curves  are taken over from Table II of \cite{Ho:2014pta}. These pairing gaps have been already used in our calculations \cite{Grigorian:2016leu}, but now we exploit EoS of the MKVORH$\phi$ model rather than the DD2 one. The corresponding zero-temperature gaps are shown on the left panel  in Fig. \ref{Protongaps}.
In our nuclear medium cooling scenario we use the temperature dependence of the gaps
taken as in \cite{Yakovlev:1999sk} and the fermion phase space integrals for the emissivities of the MMU and  MNB and DU processes and specific heat are corrected by the corresponding $R$ factors.

For $n\leq (0.5 - 0.8)\,n_0$ in a region of the inner crust there may exist a pasta phase \cite{Maruyama:2005vb}. At present the presence of the pasta is not yet included in the actual simulations of the NS cooling. At very low densities there is the outer crust and the envelope. Influence of these regions on the cooling and heat transport is minor, because of their low mass content.
Thus, the temperature changes only slightly in the region from the crust to the envelope.
In our scenario, where the cooling of the interior is strongly enhanced owing to the pion softening effect the role of the crust is still diminished.  Thereby, simplifying consideration for $n<n_{c1}$ for the MMU, MNB and the PBF emissivities we use  the expressions not including the medium effects and we disregard other possible neutrino processes yielding only tiny  contribution to the total luminosity.

As in our previous works, we use a fixed relation between the surface ($T_s$) and internal ($T$) temperatures
(see the curve ``our fit" in Fig.~4 of \cite{Blaschke:2004vq}) within the band computed in \cite{Yakovlev:2003ed},  demonstrating a similar trend as the well known  ``Tsuruta law" \cite{Tsuruta:1979}.
Our fit $T_s - T$ relation qualitatively takes into account that hotter and younger objects may have less heavy  elements in  the atmosphere than colder and older ones. A dependence of the cooling history on the choice  of the relation between  $T_s$ and $T$ was studied  in \cite{Blaschke:2004vq,Grigorian:2005fn}.

The slopes  of the cooling curves for young objects, as Cassiopeia A, are sensitive to the values of the heat conductivity of the NS matter. We use the same lepton heat conductivity as in \cite{Shternin:2007ee}, cf. \cite{Blaschke:2013vma,Grigorian:2016leu}, whereas the nucleon heat conductivity that we exploit is smaller than that used in the minimal cooling scenario due to the included pion softening effect, cf. \cite{Blaschke:2004vq,Grigorian:2005fn,Blaschke:2013vma,Grigorian:2016leu}.


On the right panel in Fig. \ref{Protongaps} we show the baryon density dependence of the $\Lambda$ hyperon pairing gaps for the MKVORH$\phi$ EoS at $T=0$. The values of the $\Lambda$ pairing gaps as functions of the $\Lambda$ Fermi momentum are taken from the calculations~\cite{TT00,TN06}.
The model TT1 uses the ND-soft model by the Nijmegen group for bare $\Lambda\Lambda$ interaction and model TTGm uses results of the G-matrix calculations by Lanskoy and Yamamoto~\cite{LanskYam}. The other three models include three-nucleon TNI6u forces for several $\Lambda\Lambda$ pairing potentials: ND-Soft, Ehime and FG-A.
Gaps are fitted with the formula proposed in~\cite{Kaminker01}. In presence of the $\Lambda\Lambda$ pairing the emissivities of the DU processes with participation of $\Lambda$ hyperons are suppressed by the $R$ factors. We use the same $R$ factors that have been derived  in \cite{Yakovlev:1999sk} for the DU processes on nucleons (see Eqs. (28)-(33) there), now with $\Lambda$ gap instead of the neutron one. The values of $\Xi^-$  gaps are poorly known.  Thereby for simplicity we consider
$\Xi^-$ unpaired.

\section{ Cooling history of neutron stars. Numerical results.}
\label{sect::results}

Below we  present results of our numerical calculations of the cooling history of NSs on the plane $T_s^{\infty}-t$, $T_s^{\infty}$ is the red-shifted surface temperature.
In figures we show cooling curves  for NSs with $M> 1\, M_{\odot}$, since with the  standard mechanisms of the formation of NSs the latter have masses $M> 1\,M_{\odot}$, and the lowest measured mass of the NS, J0453+1559, in the binary system is $M\simeq 1.17\,  M_{\odot}> 1\,M_{\odot}$, cf. table 1 in \cite{Ozel:2016oaf}.

\begin{figure}
	\centerline{
		\includegraphics[height=7.1cm,clip=true]{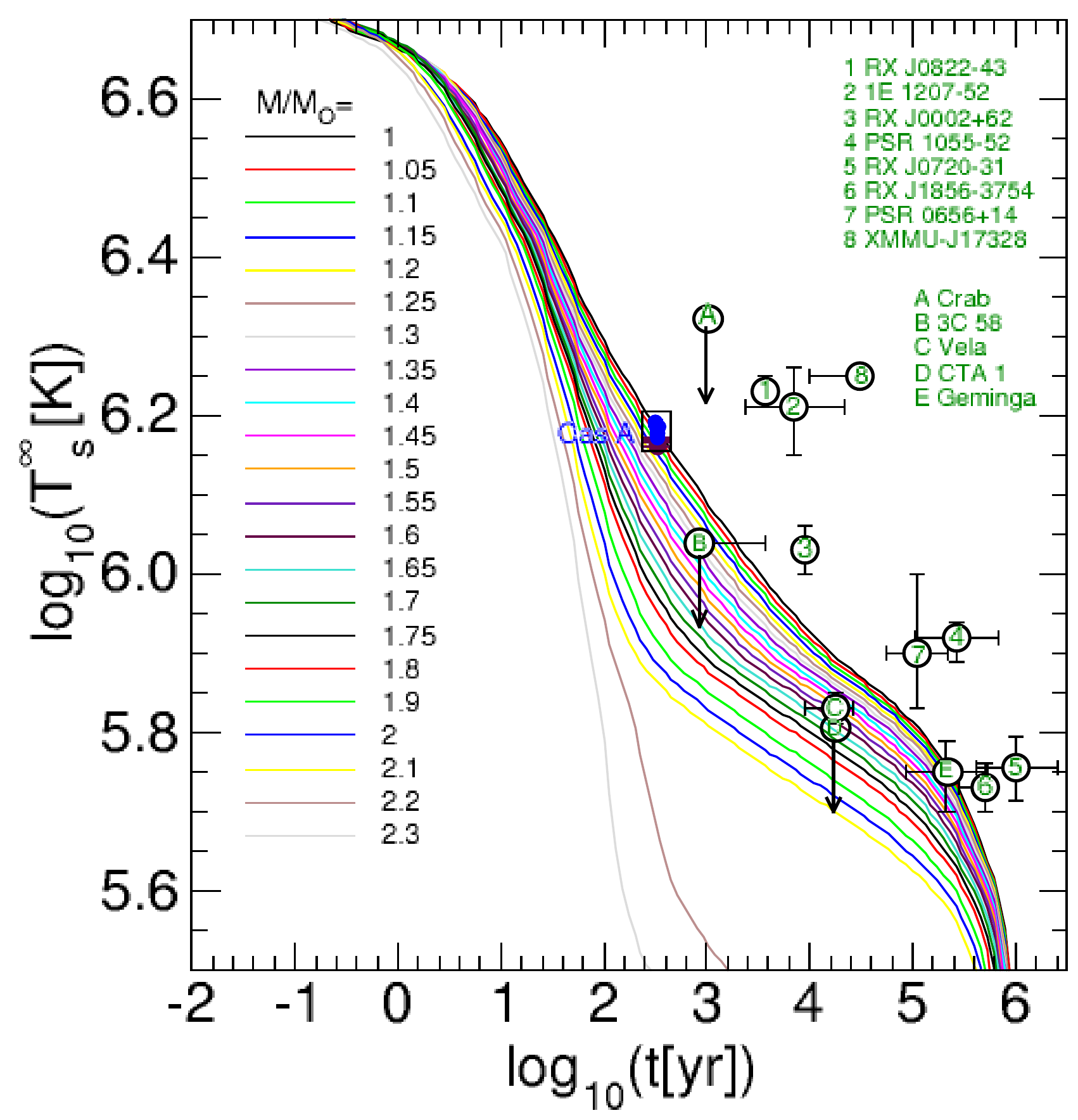}\quad
		\includegraphics[height=7.1cm,clip=true]{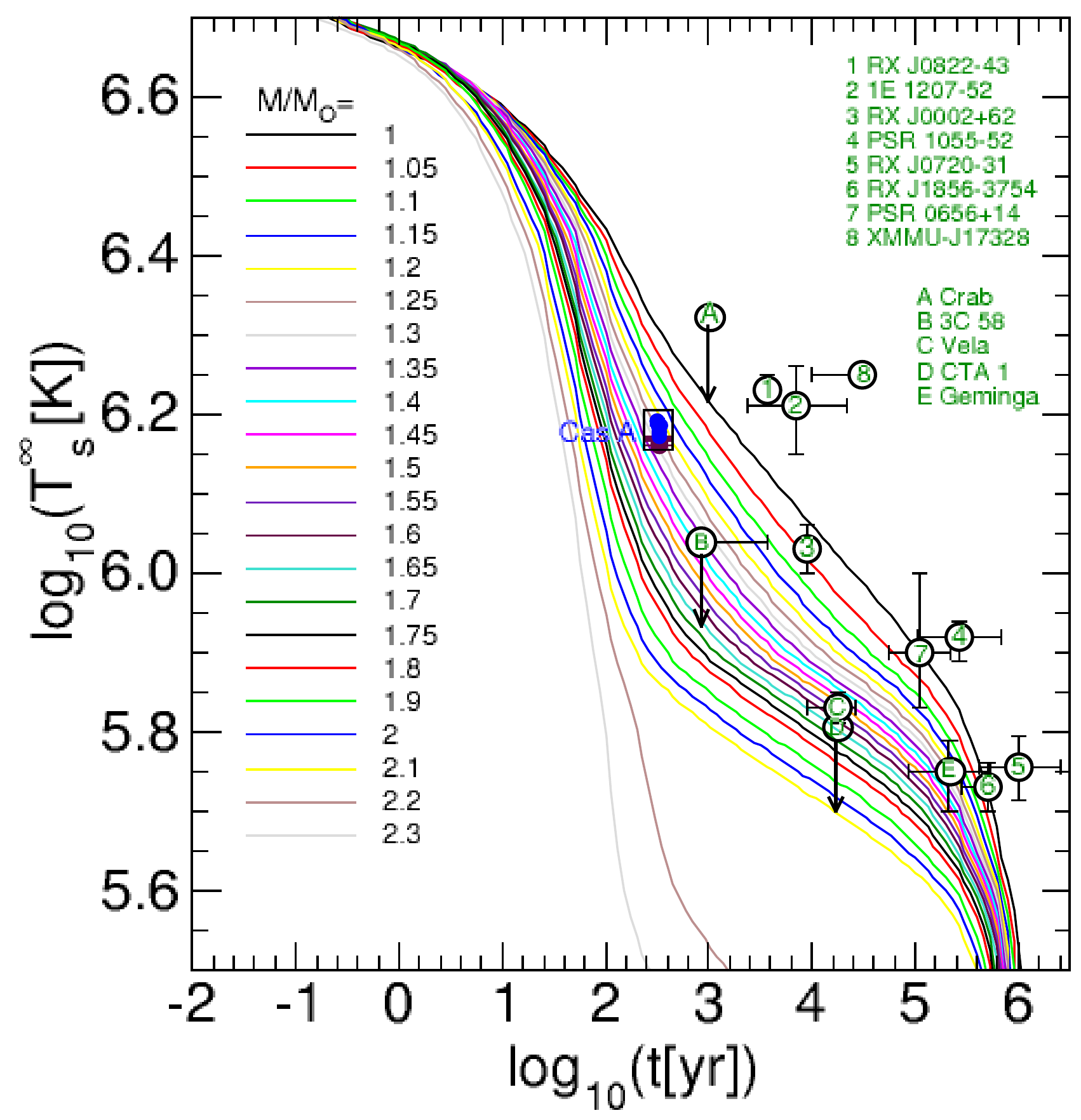}}
	\caption{ Redshifted surface temperature as a function of the NS age for various NS masses for MKVOR model (without hyperons). Effective pion gap follows the solid curve in Fig. \ref{piongap}. Left panel:  Proton gaps  are taken following BS model. Right panel:  Proton gaps  are taken following CCYps model.  }
	\label{Nohyperon1}
\end{figure}

\begin{figure}
	\includegraphics[height=7.1cm,clip=true]{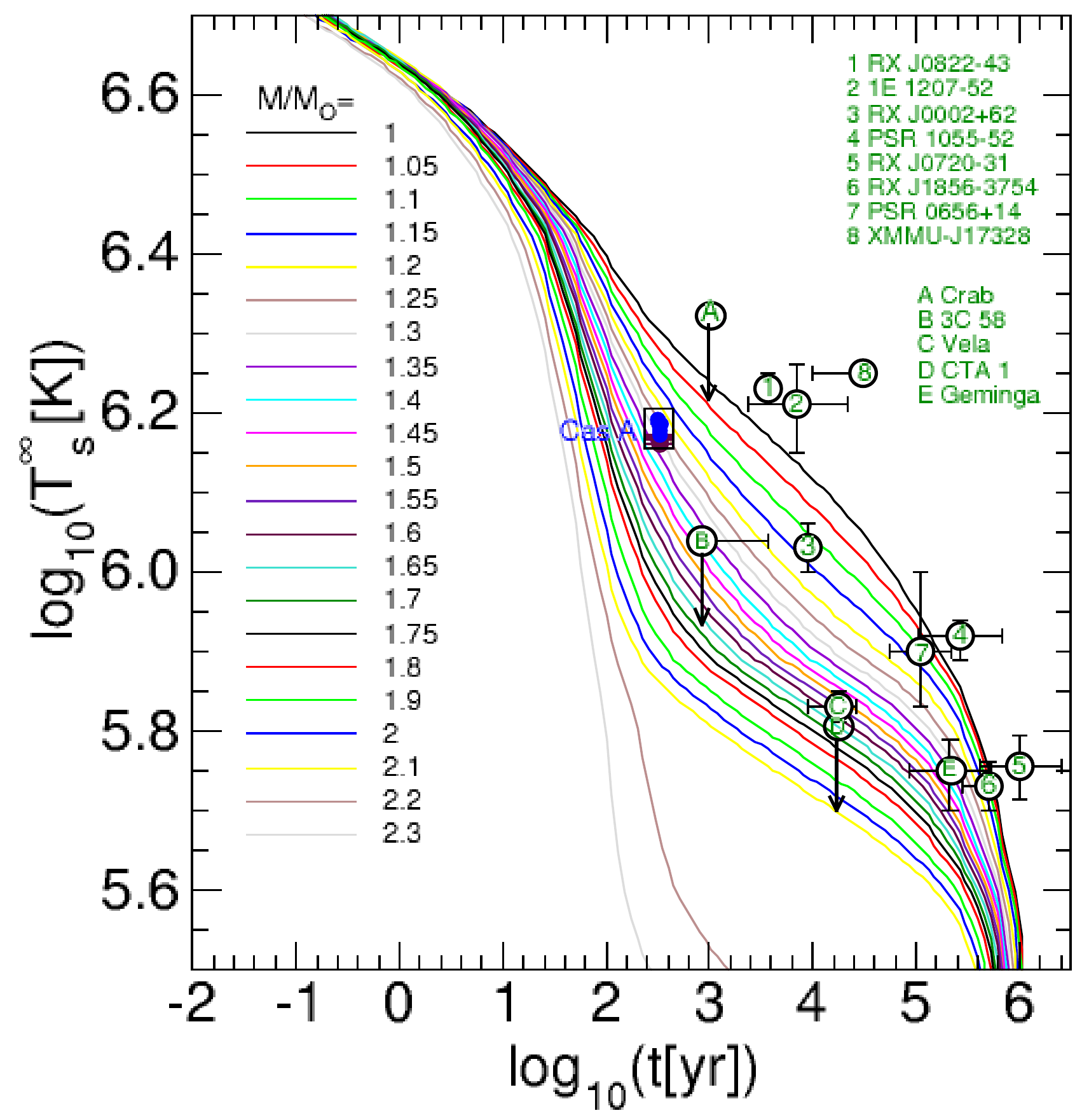}\quad
	\includegraphics[height=7.1cm,clip=true]{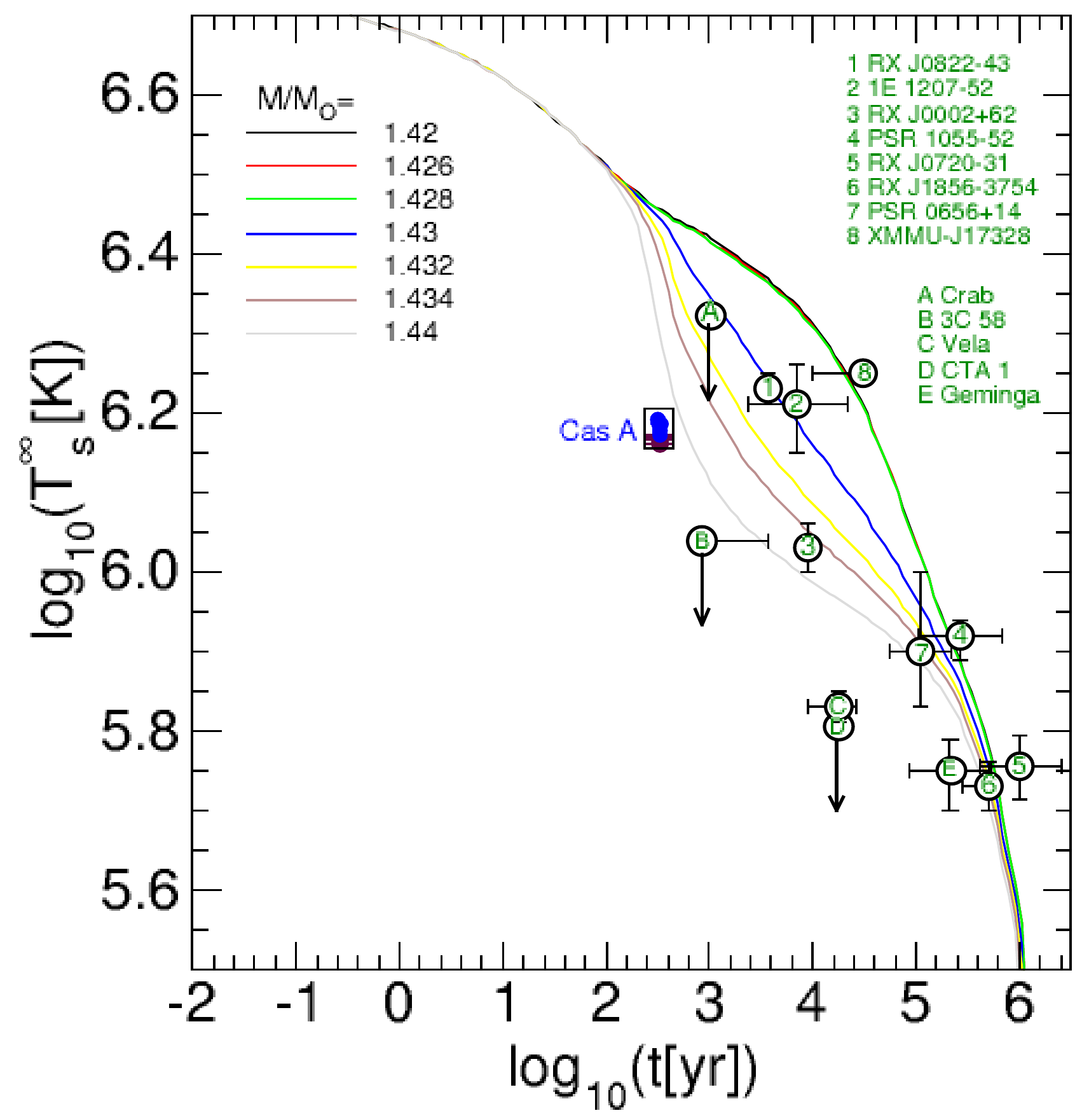}
	\caption{ Redshifted surface temperature as a function of the NS age for various NS masses. Left panel: MKVOR model without hyperons. Effective pion gap follows the solid curve in Fig. \ref{piongap}. Proton gaps  are taken following AO model. Right panel: MKVORH$\phi$ model with hyperons. The proton gaps are chosen following the EEHO model. The mass set demonstrates the effect of the DU process on $\Lambda$ hyperons with the threshold mass of 1.429~$M_\odot$.}
	\label{Nohyperon2}
\end{figure}

In Figs. \ref{Nohyperon1} (left and right panels) and in Fig. \ref{Nohyperon2} (left panel) we show the cooling history of NSs  calculated using the EoS of the MKVOR model without inclusion of hyperons, with the proton pairing gaps   taken following the models BS, CCYps and AO, respectively. We used the effective pion gap given by the solid curve  in Figure \ref{piongap}, $n_c^{\rm PU}=3 \, n_0$. The BS proton gap   vanishes  for $n>1.852 \, n_0$. Note that  already for $\Delta_p <(0.01-0.02)\,{\rm MeV}$  for internal temperatures  $10^8-10^9\,{\rm K}$ of our interest the effect of the suppression of the  neutrino emissivity given by the $R<1$ factor becomes not  significant.
The density
$n_{\rm cen}\geq 1.852 \, n_0$ is reached in the center of the  NS with $M\simeq  0.585 \, M_{\odot}$, see Fig. \ref{mkv_mass} (right). Thereby in the given MKVOR  model with BS proton gaps in NSs with higher  masses, having  in a broad interior regions $n> 1.852 \, n_0$,  the protons can be considered as not paired.
The CCYps proton gap  vaishes for $n> 2.337 \, n_0$. The value $n_{\rm cen}=2.337 \, n_0$ is reached for $M\simeq  1.188\, M_{\odot}$.  The AO gap vanishes for $n>2.640 \, n_0$, corresponding to  $M = 1.443 \, M_{\odot}$. However already for $M>1.1 \, M_{\odot}$, when the central density in this model exceeds  $n_{\rm cen} \simeq 2.321 \, n_0$,  the proton gap becomes less than $0.01 $MeV.
Thereby, in the models exploiting BS, CCYps and AO proton gaps    the MMU processes involving protons prove to be very efficient  in all NSs with $M> (1-1.2) \, M_{\odot}$ (since protons in broad central regions of these NSs   have very small or even zero pairing gaps). Due to that, as we see in  Figs. \ref{Nohyperon1} left and right
and  in Fig. \ref{Nohyperon2}, the cooling curves lie too low to appropriately explain surface temperatures  for  the group of the NSs being the slow coolers. If we used a smaller effective pion gap, e.g. corresponding to $n_c^{\rm PU}=2.5 \, n_0, 2 \, n_0$ and $1.5 \, n_0$, as shown by the dashed, dotted or dash-dotted curves in Figure \ref{piongap}, the cooling curves calculated with BS, CCYps and AO proton pairing gaps would lie still lower. Thus we conclude that within our nuclear medium cooling scenario the $T_s^{\infty} - t$ data cannot be explained using the models BS, CCYps and AO for the proton gaps. The difference of the cooling curves computed with the MKVOR model without hyperons and the MKVORH$\phi$ model with hyperons arises only for $n>2.621 \, n_0$, $M > 1.426 \, M_{\odot}$ when $\Lambda$ baryons start to appear in the central regions of the NSs. Thus, the density dependence of  the proton gaps in BS, CCYps   models  is the same in the whole available density interval for both the MKVOR and MKVORH$\phi$ EoSs, and for the AO model hyperons may coexist with superfluid protons  only in a narrow central regions, where $2.621 \, n_0<n<2.64 \, n_0$. The DU process on $\Lambda$-hyperons occurs for $n> 2.625 \, n_0$ that translates to the NS mass $M>M_{c,\Lambda}^{\rm DU}\simeq 1.429 \, M_{\odot}$. With the allowed DU process on $\Lambda$   the NS cooling is strongly accelerated. Thereby with BS, CCYps and AO proton pairing gaps we cannot explain the $T_s (t)$ data in both the MKVOR and MKVORH$\phi$ models.

On the right panel of the Fig. \ref{Nohyperon2} we demonstrate the effect of the $\Lambda\to p + e + \bar \nu$ DU process on the cooling curves by performing a small variation of the NS mass near the threshold mass for the process $M_{c \Lambda}^{\rm DU} \simeq  1.429\, M_\odot$. For the demonstrative purposes we chose here the EEHO proton gap model. We see that for masses in the interval $1.42 - 1.428\, M_\odot < M_{c \Lambda}^{\rm DU}$ the cooling curves are  insensitive to  a small variation of the source's mass and almost indistinguishable one from another. However, for $M > M_{c \Lambda}^{\rm DU}$ the contribution of the DU process on $\Lambda$ makes the cooling curves very sensitive to the object's mass. Thereby the curve for $1.430\, M_\odot > M_{c \Lambda}^{\rm DU} $ lies substantially lower than that for $1.428\, M_\odot < M_{c \Lambda}^{\rm DU} $.

\begin{figure}
	\centerline{\includegraphics[height=7.1cm,clip=true]{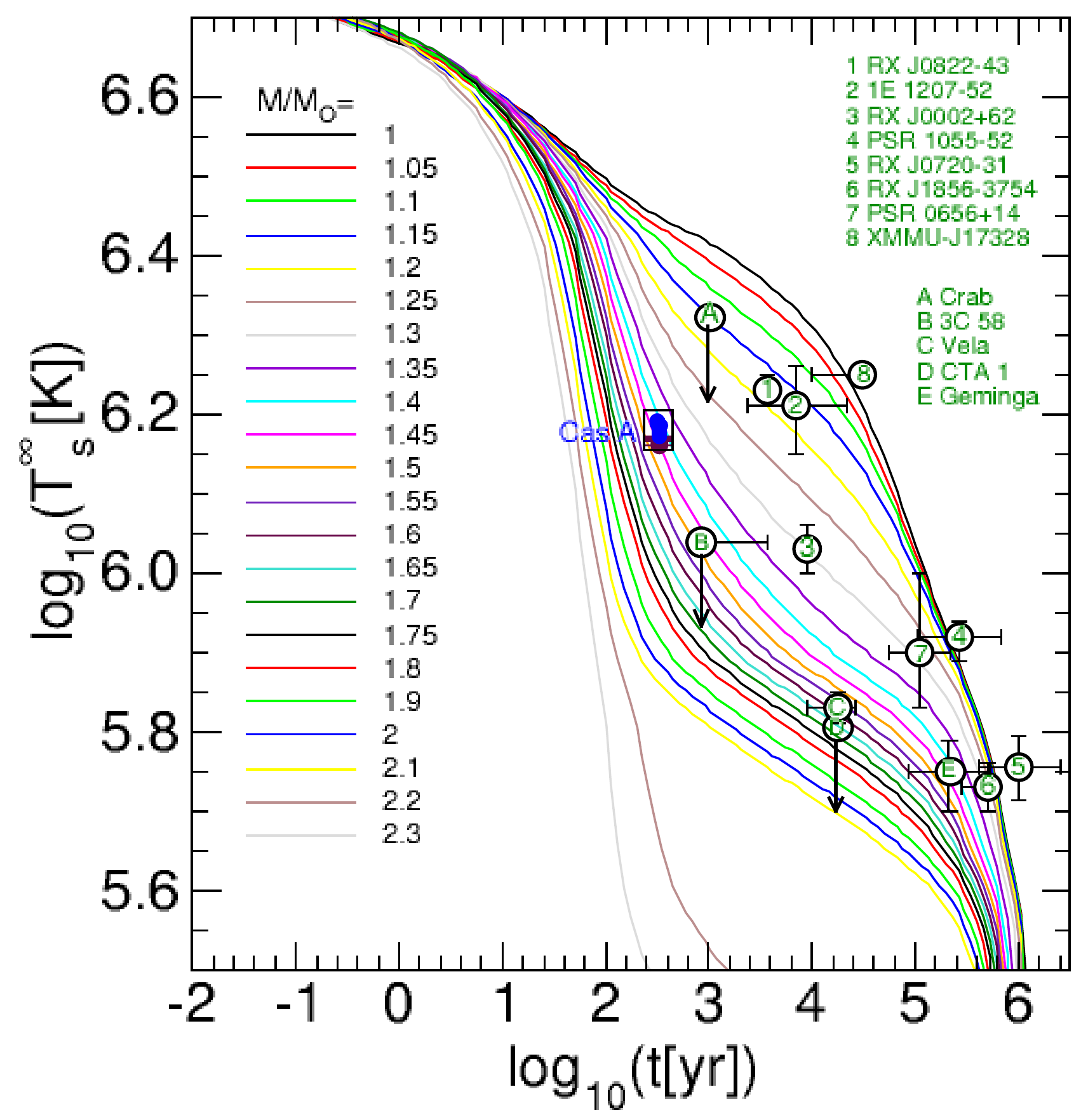}\quad
		\includegraphics[height=7.1cm,clip=true]{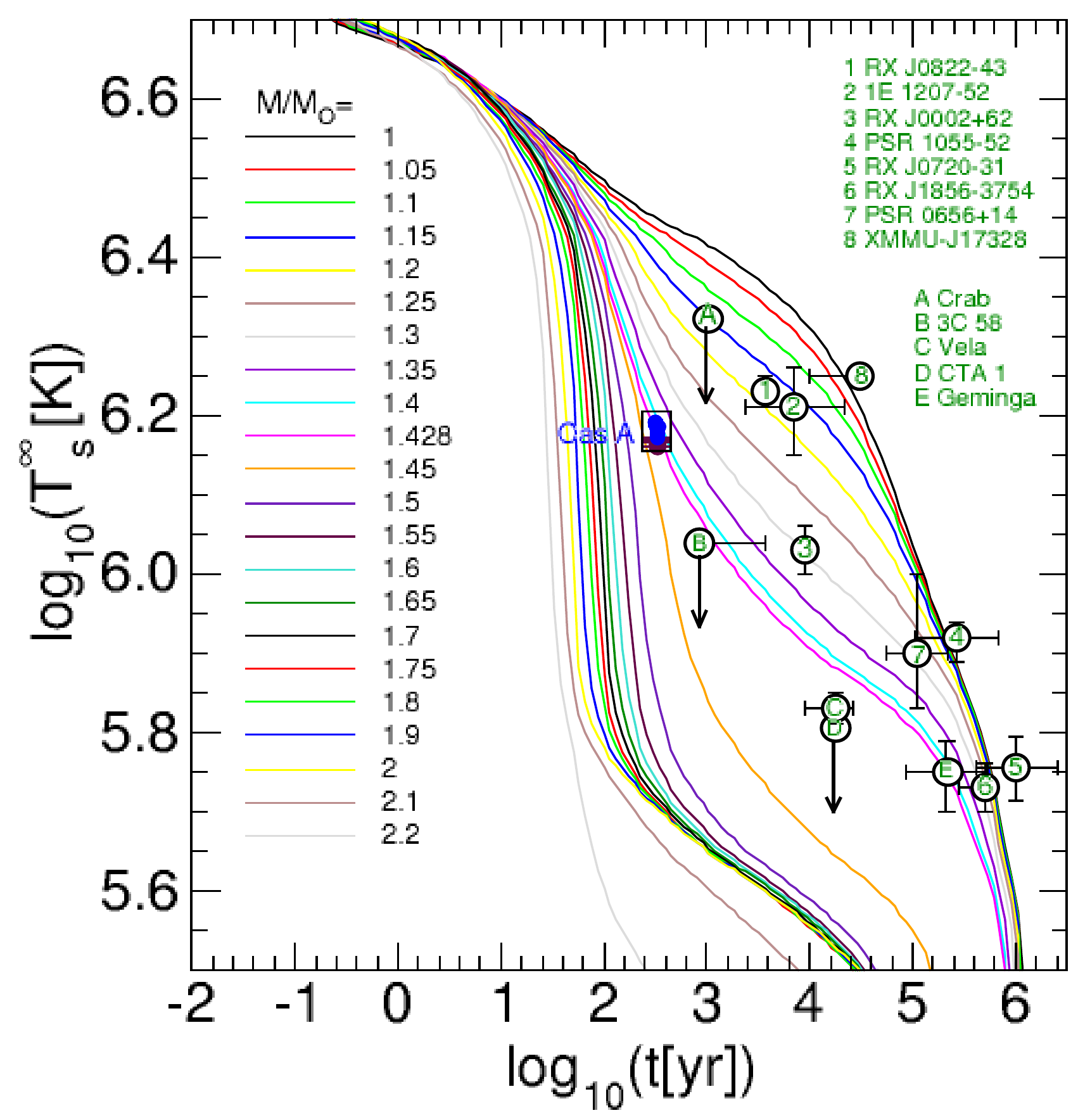}}
	\caption{ Redshifted surface temperature as a function of the NS age for various NS masses. Proton gaps  are taken following BCLL model.
		Effective pion gap follows the solid curve in Fig. \ref{piongap}. Left panel: MKVOR model (without hyperons).  Right panel: MKVORH model (with hyperons). Hyperon gaps are given by the TN-FGA
		model.}
	\label{NoHhyperon3}
\end{figure}

\begin{figure}
	\centerline{
		\includegraphics[height=7.1cm,clip=true]{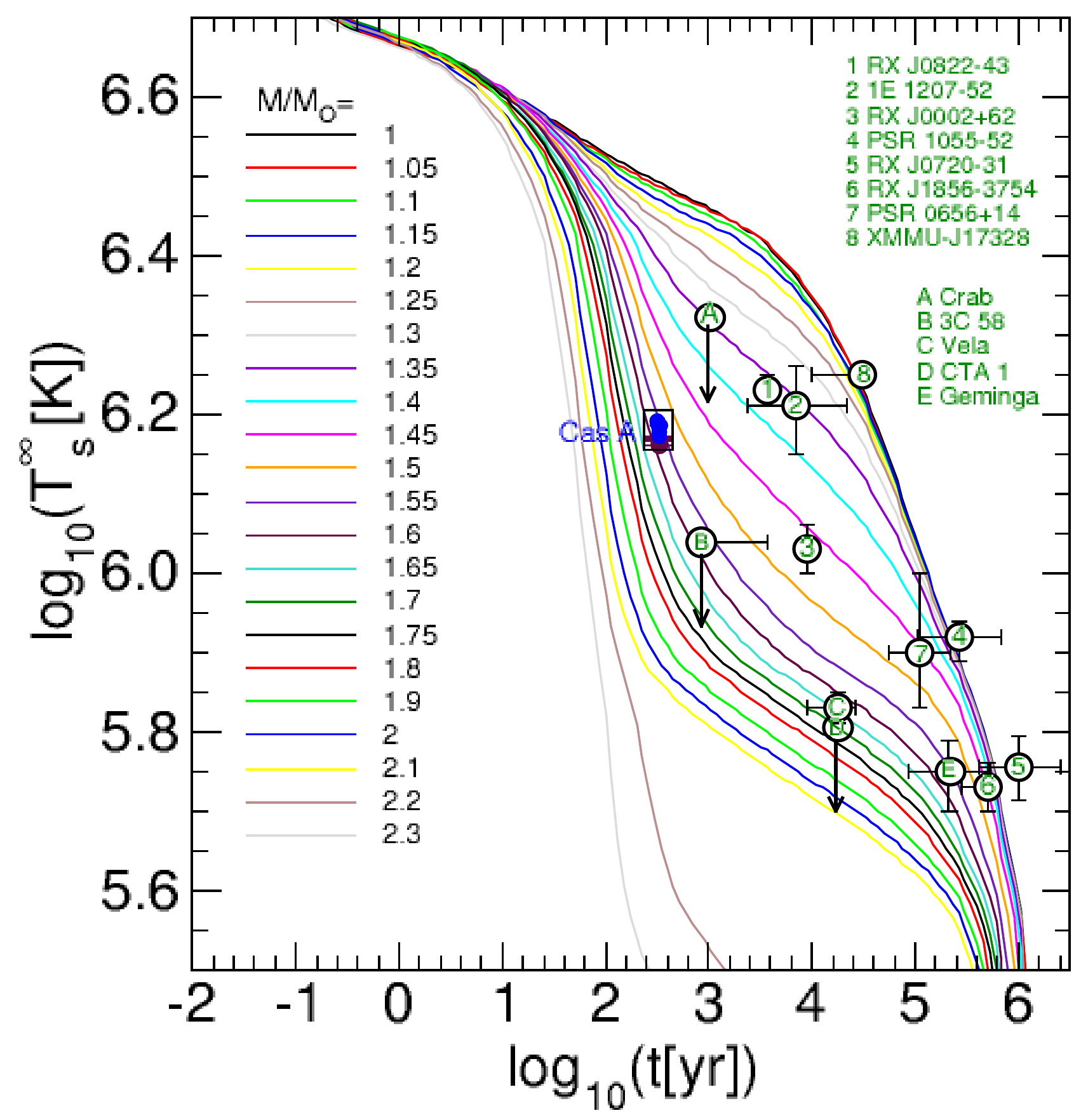}\quad
		\includegraphics[height=7.1cm,clip=true]{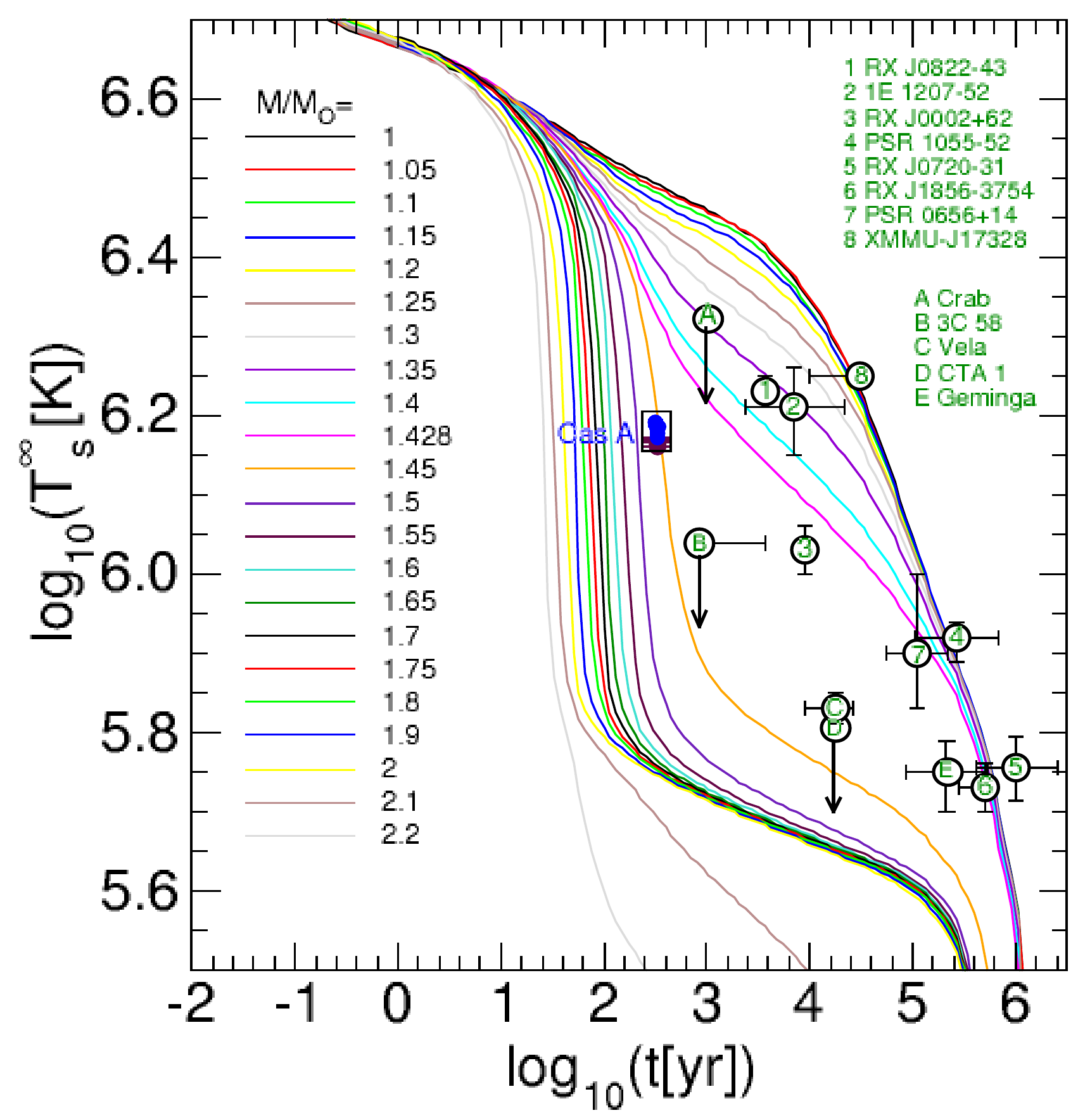}}
	\caption{ Redshifted surface temperature as a function of the NS age for various NS masses. Proton gaps  are taken following CCYms model.
		Effective pion gap follows the solid curve in Fig. \ref{piongap}. Left panel: MKVOR model (without hyperons).  Right panel: MKVORH model (with hyperons). Hyperon gaps follow the TN-FGA model.}
	\label{NoHhyperon4}
\end{figure}

\begin{figure}
	\centerline{\includegraphics[height=7.1cm,clip=true]{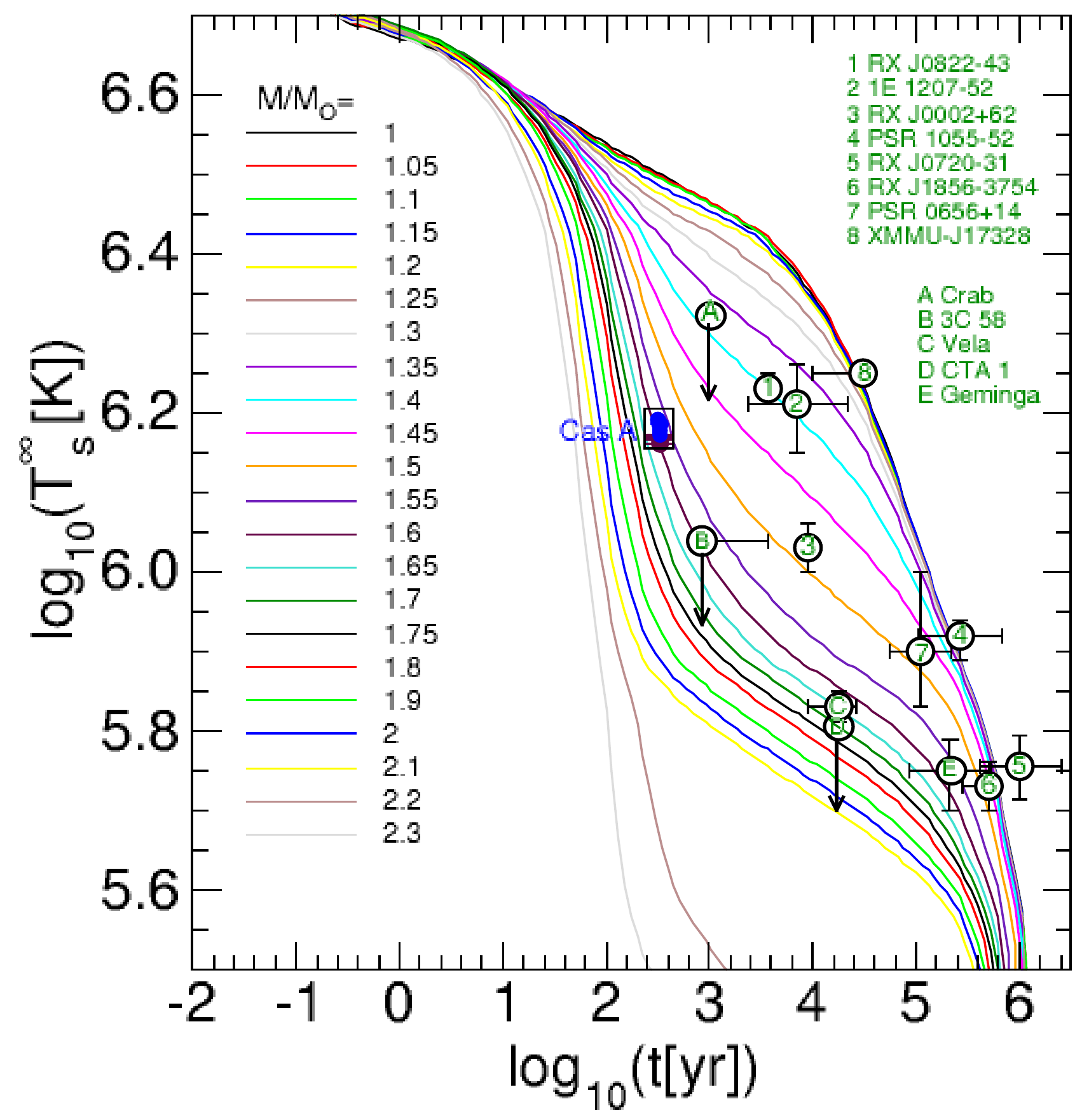}\quad
		\includegraphics[height=7.1cm,clip=true]{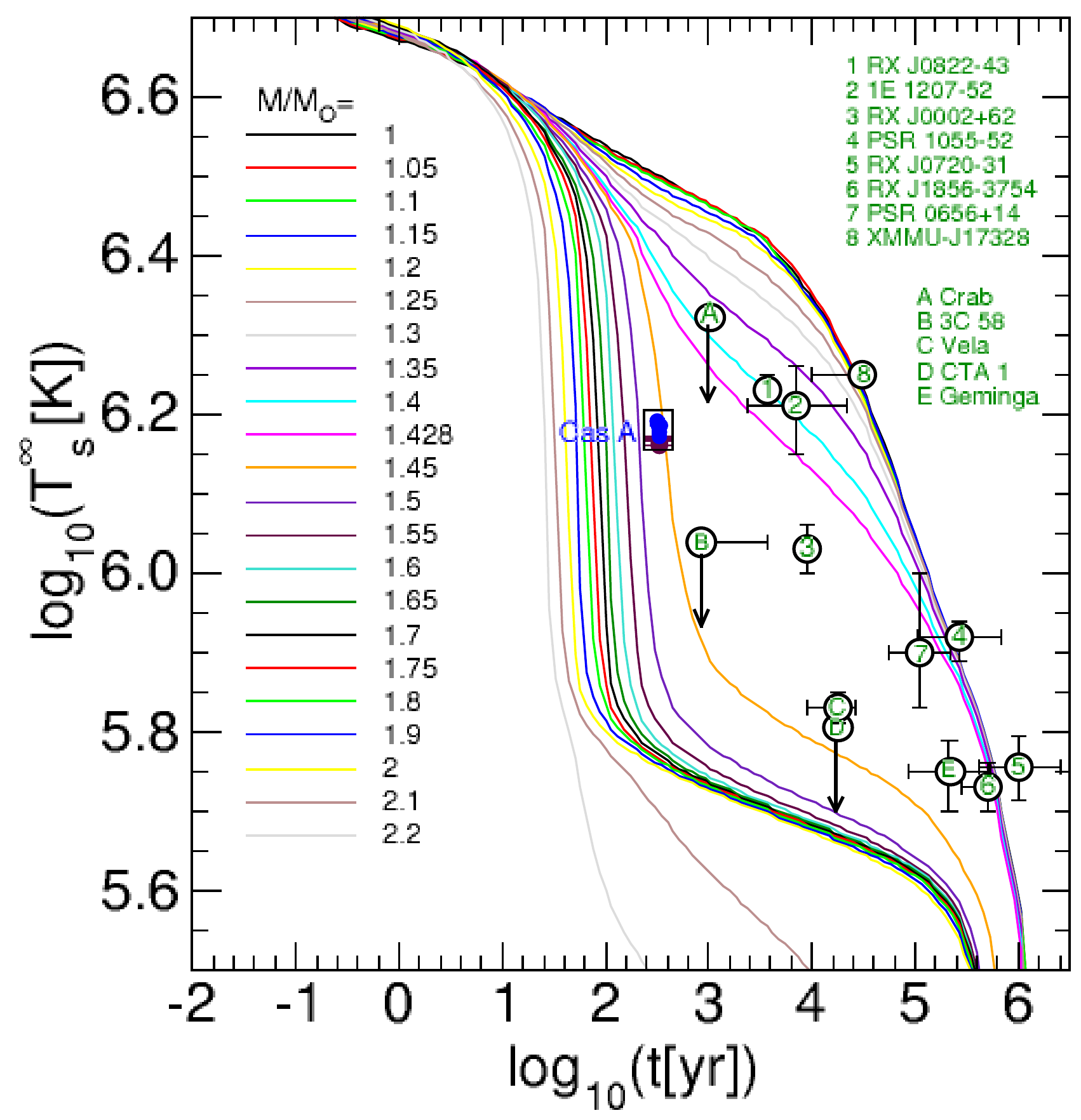}}
	\caption{ Redshifted surface temperature as a function of the NS age for various NS masses. Proton gaps  are taken following EEHOr model. Effective pion gap follows the solid curve in Fig. \ref{piongap}. Left panel: MKVOR model (without hyperons).  Right panel: MKVORH model (with hyperons). Hyperon gaps follow TN-FGA model. }
	\label{NoHhyperon5}
\end{figure}

\begin{figure}
	\centerline{
		\includegraphics[height=7.1cm,clip=true]{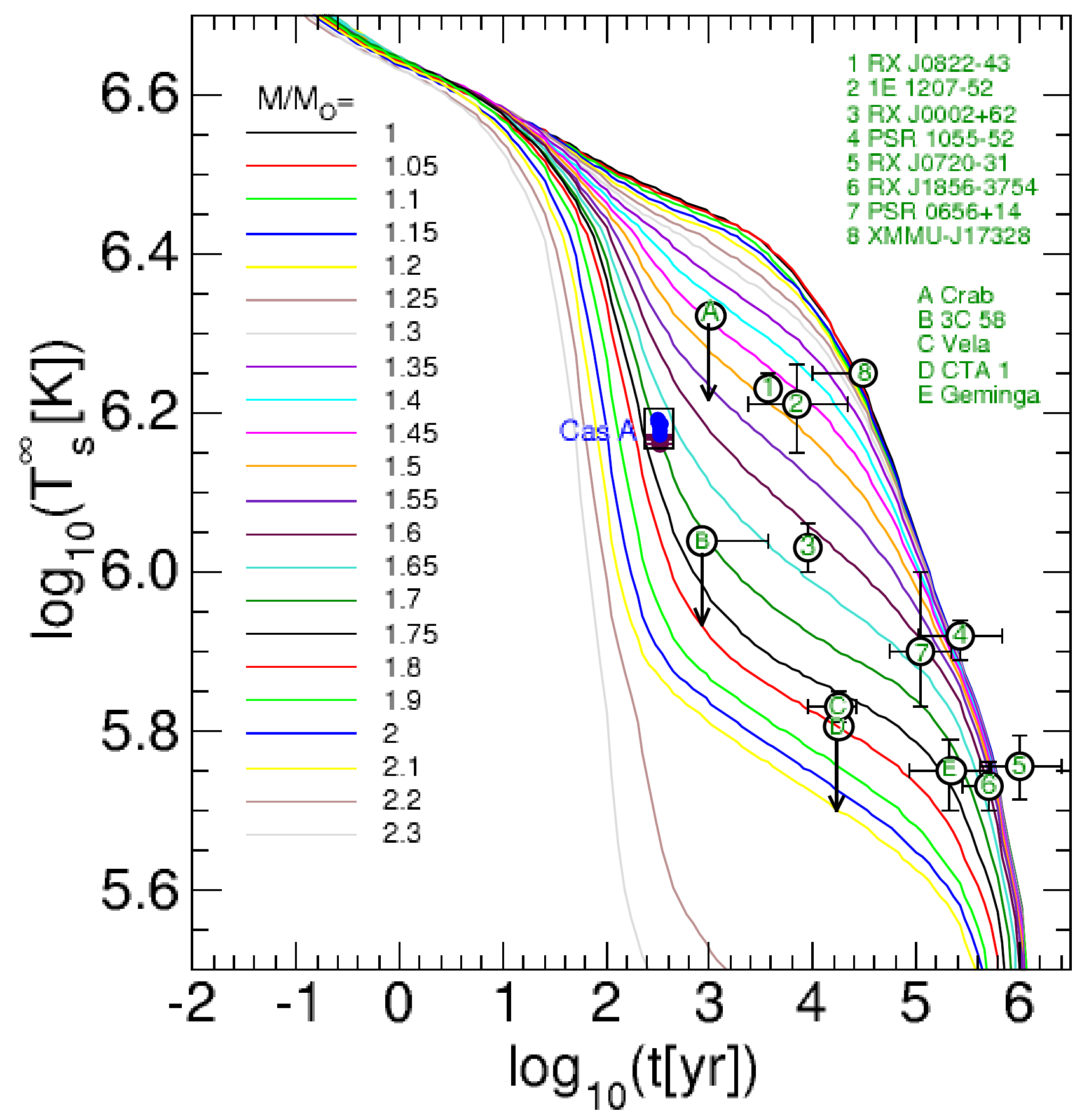}\quad
		\includegraphics[height=7.1cm,clip=true]{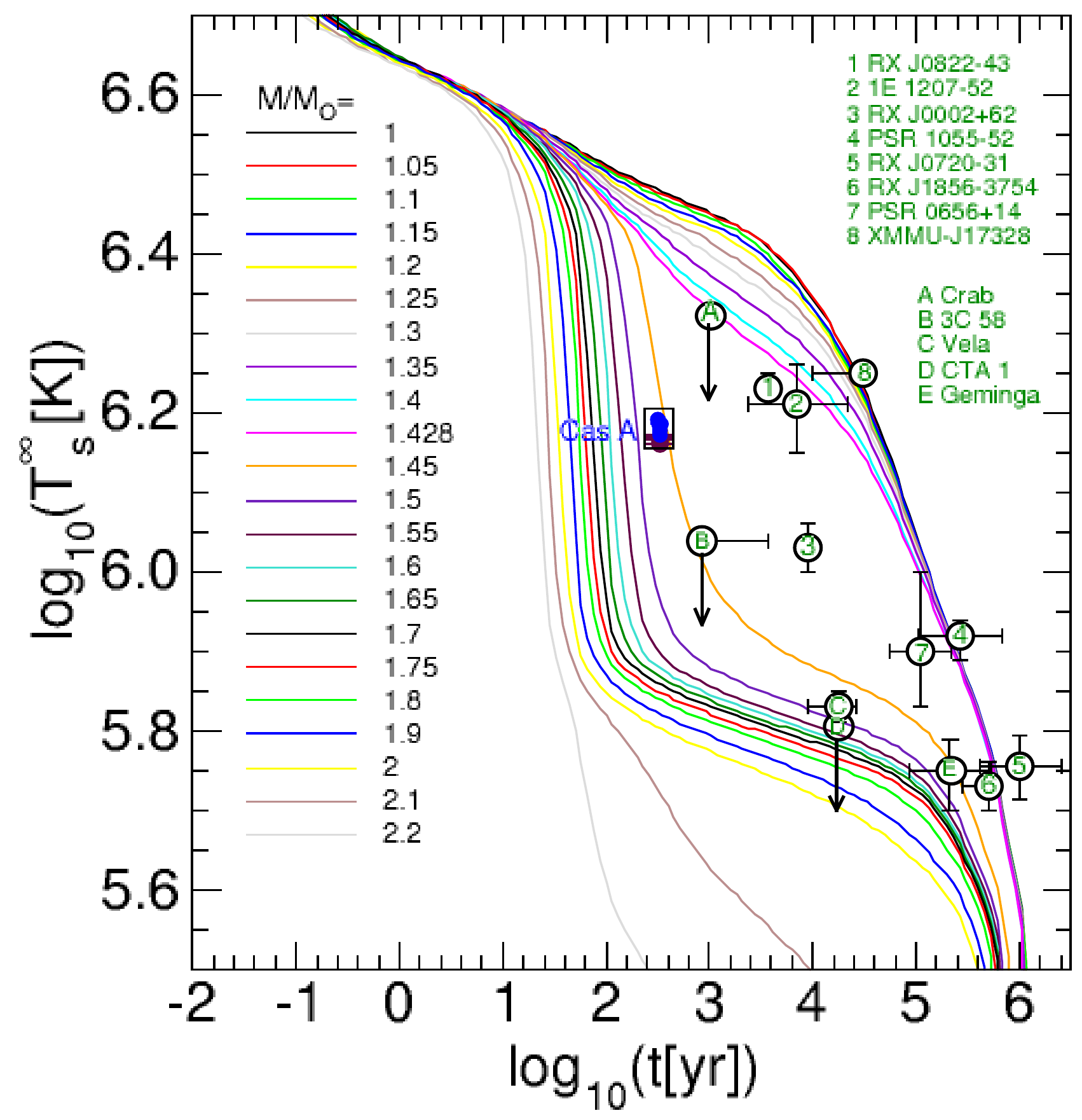}}
	\caption{ Redshifted surface temperature as a function of the NS age for various NS masses. Proton gaps  are taken following T model. Effective pion gap follows the solid curve in Fig. \ref{piongap}. Left panel: MKVOR model (without hyperons).  Right panel: MKVORH model (with hyperons). Hyperon gaps follow TN-FGA model. }
	\label{NoHhyperon6}
\end{figure}

\begin{figure}
	\centerline{\includegraphics[width=0.45\textwidth]{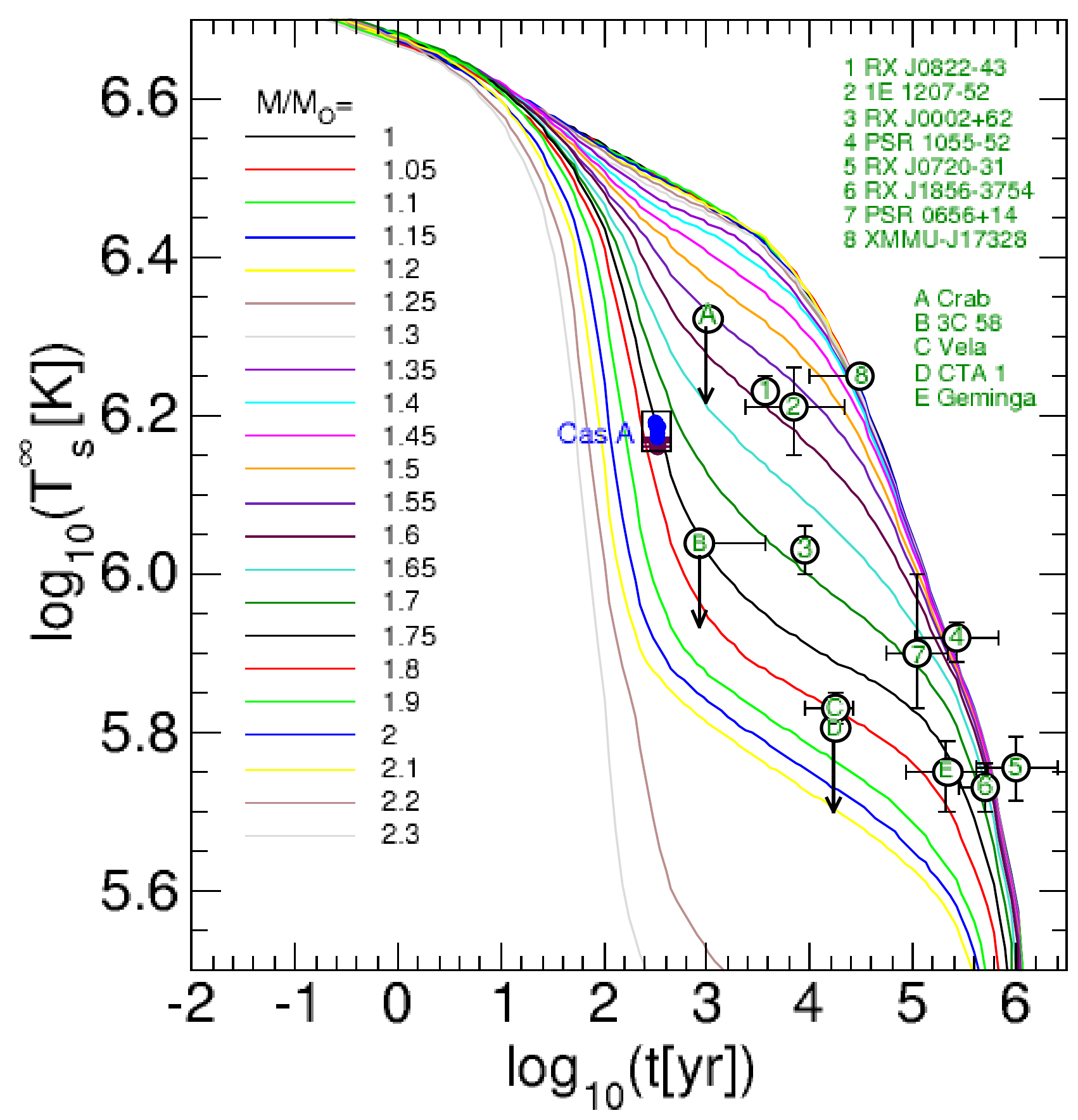}\quad
	\includegraphics[width=0.45\textwidth]{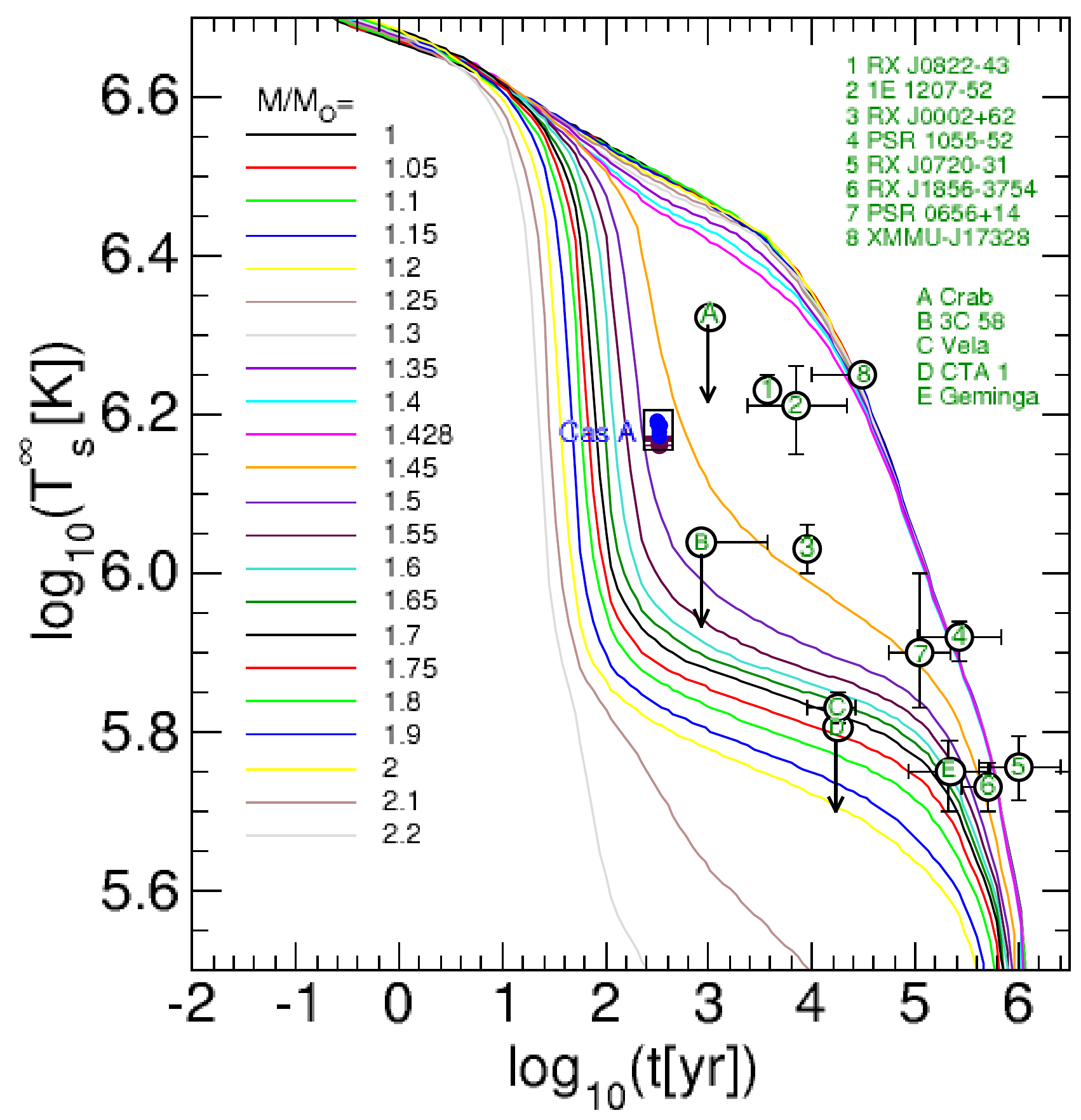}}
	\caption{ Redshifted surface temperature as a function of the NS age for various NS masses. Proton gaps  are taken following EEHO model. Effective pion gap follows the solid curve in Fig. \ref{piongap}. Left panel: MKVOR model (without hyperons).  Right panel: MKVORH model (with hyperons). Hyperon gaps follow TN-FGA model. }
	\label{NoHhyperon7}
\end{figure}

\begin{figure}
	\centerline{\includegraphics[height=7.1cm,clip=true]{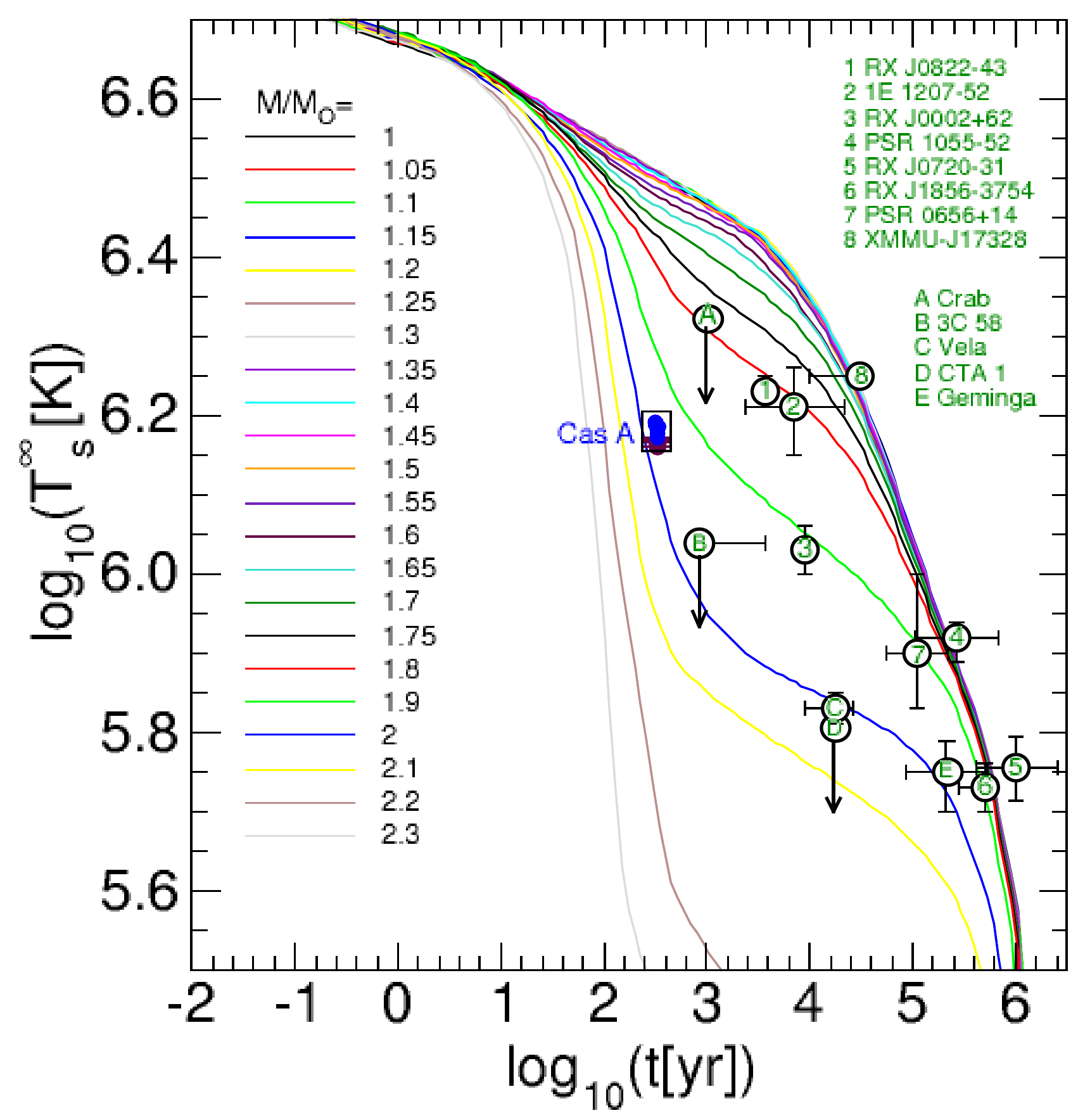}\quad
		\includegraphics[height=7.1cm,clip=true]{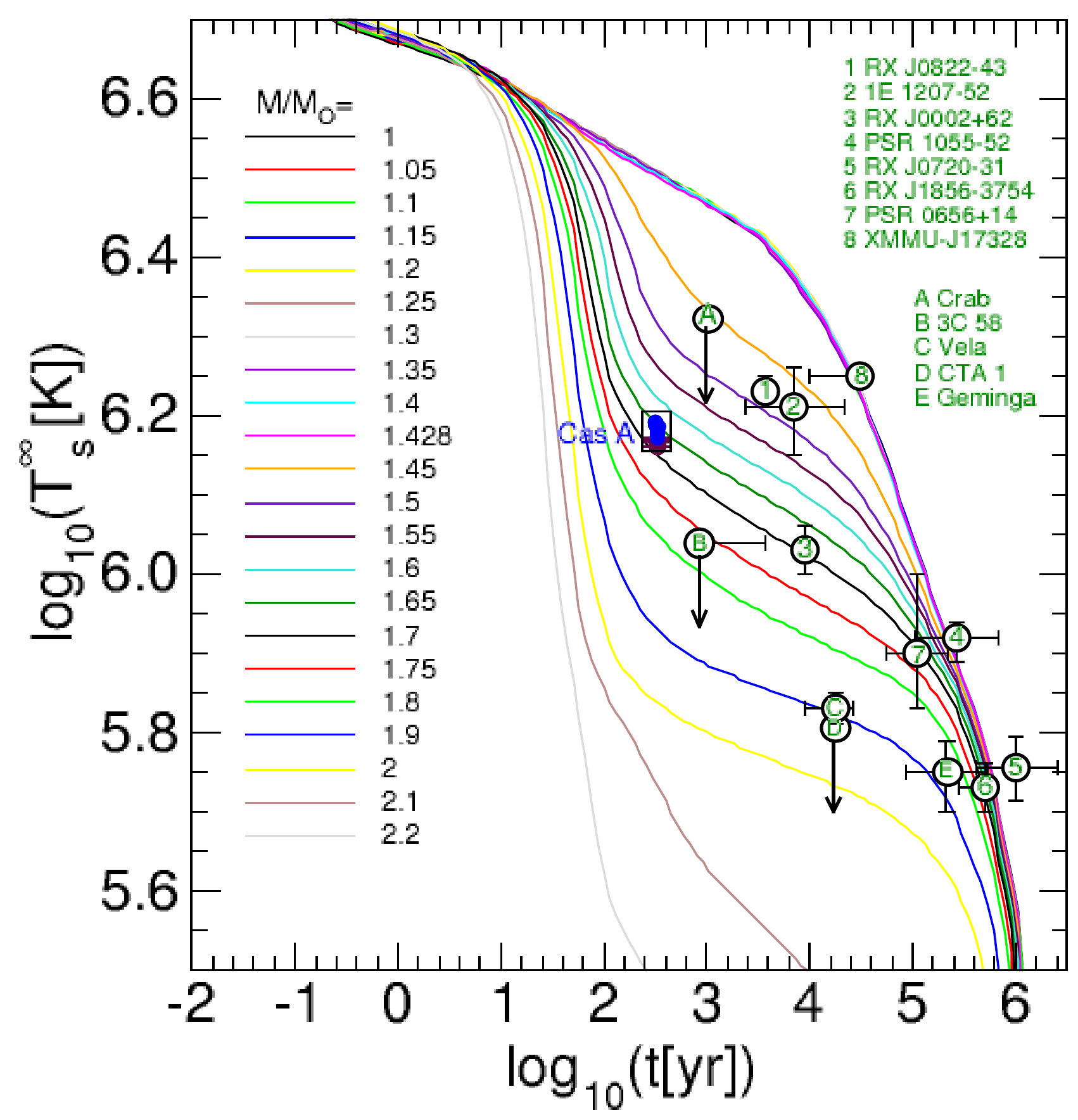}}
	\caption{ Redshifted surface temperature as a function of the NS age for various NS masses. Proton gaps  are taken following CCDK model. Effective pion gap follows the solid curve in Fig. \ref{piongap}. Left panel: MKVOR model (without hyperons).  Right panel: MKVORH model (with hyperons). Hyperon gaps follow TN-FGA model. }
	\label{NoHhyperon8}
\end{figure}

In  Figs. \ref{NoHhyperon3} - \ref{NoHhyperon8}
we show the cooling history of NSs calculated using the EoS of the MKVOR model without inclusion of hyperons (in left panels), and  the  results for the MKVORH$\phi$ EoS with inclusion of hyperons (in right panels).  Proton gaps are taken following the models  BCLL, CCYms, EEHOr, T, EEHO and CCDK in these figures, which are shown in Fig. \ref{Protongaps} (left) for the MKVORH$\phi$ EoS. Effective pion gap   follows the solid curve in Fig. \ref{piongap}. In case of MKVORH$\phi$ EoS (on right panels in Figs. \ref{NoHhyperon3} - \ref{NoHhyperon8})  we use hyperon gaps  given by the TN-FGA
model, shown in Fig. \ref{Protongaps}, right.

In Fig. \ref{NoHhyperon3} (left and right)  we exploit proton pairing gaps following the BCLL model.  As we see in the left panel, in the MKVOR model with BCLL proton gaps the objects 8, 4, 5, A, 1, 2 form a family of slow coolers.
The slow coolers have masses in the interval $M=(1-1.2)M_{\odot}$. The objects 3, 7, 6, E, Cas A in this model  form a family of intermediate coolers.
The NS in Cas A    has the mass varying in the interval $M\simeq (1.4-1.45) \, M_{\odot}$. The rapid coolers B, C, D in this model are described by objects with masses  $M= (1.5-2) \, M_{\odot}$. The dependence of the cooling curves on the NS mass is very regular in the broad mass interval $1 \, M_{\odot}<M_c^{\rm DU} \simeq 2.149 \, M_\odot$. The main cooling regulator for $M<M_c^{\rm DU}$ is the MMU process.  For $M > M_c^{\rm DU}$ there appears very efficient DU process on nucleons. Thereby the NSs with $M > M_c^{\rm DU}$ cool down very rapidly by the DU processes on nucleons.

In the MKVORH$\phi$ model the proton BCLL gap  vanishes for $n > 2.646 \, n_0$.   The central density $n_{\rm cen}=2.646\, n_0$ is reached   in the MKVORH$\phi$ model for the mass $M = 1.449 \, M_{\odot}> M_{c\Lambda}^{\rm DU}$. In the MKVORH$\phi$ model
$\Lambda$ hyperons appear first  for $M=1.421 \,  M_{\odot}$. As we see from the figure, the hyperons do not exist in the NSs, which are slow  and intermediate coolers in this model. The Cas A object is described as having the mass $M\simeq (1.4-1.428) \, M_{\odot}$ (no DU on $\Lambda$ yet). Only cooling history of the objects B, C, D is determined by the  rapid $\Lambda$ hyperon DU processes in this model.   Masses of the rapid coolers, B, C, D, vary in a narrow  interval $M\simeq (1.429-1.45)\, M_{\odot}$. In this narrow mass interval the value of the $\Lambda$ pairing gap is still rather small and the $\Lambda$ hyperon DU process is not yet significantly suppressed by the corresponding $\Lambda$ superfluidity factor. For $M>2.078 \, M_{\odot}$ there appear very efficient DU processes on nucleons and thereby such stars are cooled very rapidly by the DU processes on nucleons.

In Fig. \ref{NoHhyperon4} (left and right)  we show the same as in Fig. \ref{NoHhyperon3}, but now for proton pairing gaps following the   CCYms model.  As we see in the left panel,
objects 8, 4, 5, A, 1, 2 form a family of slow coolers, now with masses $M=(1-1.4)M_{\odot}$. The objects 3, 7, 6, E, Cas A, B in this model  form a family of intermediate coolers.
The NS in Cas A in MKVOR model  is appropriately described by the object with $M\simeq (1.55-1.6)\, M_{\odot}$. These values are   higher than those we obtained using the BCLL  model for the proton gaps, since  the CCYms proton gaps are typically larger than the BCLL ones. Rapid coolers C, D have masses $M\simeq (1.65-2)M_{\odot}$. The dependence of the cooling curves on the NS mass remains very regular  for $1\, M_{\odot}<M<M_c^{\rm DU} \simeq 2.149 \, M_\odot$ similar to that we obtained using the BCLL proton gaps.

In the MKVORH$\phi$ model  the CCYms proton gap  vanishes for $n>2.874 \, n_0$. The central density $n_{\rm cen}=2.874 \, n_0$ is reached  for  $M =  1.620 \, M_{\odot}$. The  hyperons do not exist in the objects 8, 4, 5, A, 1, 2, which are slow coolers in this model without hyperons in their interiors. Intermediate coolers 6, 7, 3, E,  as well as rapid coolers, Cas A and points B, C, D, are described as being cooled  mainly  by the DU processes on $\Lambda$s, and have masses in a narrow interval $1.428 \, M_{\odot}<M<(1.45-1.47) \, M_{\odot}$.  In this  narrow mass interval the value of the $\Lambda$ pairing gap remains still  small and the $\Lambda$ hyperon DU process is not yet significantly suppressed by the corresponding $\Lambda$ superfluidity $R$ factor.
The Cas A object has the mass $M \simeq (1.45-1.47) \, M_{\odot}$.

In Fig. \ref{NoHhyperon5} (left and right)  we show the same as in Fig. \ref{NoHhyperon3}, \ref{NoHhyperon4}, but  proton pairing gaps are now given by the   EEHOr model. As we see in the left panel, in the MKVOR model the
objects 8, 4, 5, A, 1, 2 form a family of slow coolers with masses $M\simeq (1-1.4) \, M_{\odot}$.
The objects 3, 7, 6, E, Cas A, B in this model  form a family of intermediate coolers.   The object in Cas A   is appropriately explained now by the NS with  the mass $M\simeq 1.6 \, M_{\odot}$. Rapid coolers C, D have masses in the interval $M\simeq (1.65-2)\,M_{\odot}$. The dependence of the cooling curves on the NS mass remains very regular for $1\, M_{\odot}<M<2.149 \, M_\odot$.

In the MKVORH$\phi$ model the EEHOr proton gap  vanishes for $n>2.875\, n_0$, very similar to that we had using the CCYms proton gaps. The central density $n=2.875 \, n_0$ corresponds to the NS mass $M=1.620 \, M_{\odot}$. However in the whole density interval the proton gaps in the EEHOr model remain larger than in the CCYms one and thereby the cooling processes in the former model are more suppressed by the corresponding $R$ factors.
The  hyperons do not exist in the objects 8, 4, 5, A, 1, 2, which are slow  coolers in this model.  Intermediate coolers 6, 7, 3, E,  as well as rapid coolers Cas A, B, C, D, are described by the cooling due to the DU processes on $\Lambda$s in a  mass interval $1.429\, M_{\odot}<M<1.48\, M_{\odot}$. Cas A object has the mass $M \simeq (1.45-1.48) \, M_{\odot}$.

In Fig. \ref{NoHhyperon6} (left and right)  we show the same as in Fig. \ref{NoHhyperon3}, \ref{NoHhyperon4}, \ref{NoHhyperon5}, but for proton pairing gaps following the   T model. As we see in the left panel,
objects 8, 4, 5, A, 1, 2 form a family of slow coolers with masses $M\simeq (1-1.5)M_{\odot}$. The objects 3, 7, 6, E, Cas A, B in this model  are intermediate coolers. As we see, the NS in Cas A in the MKVOR model with T proton gap is appropriately explained by the object with $M\simeq 1.7\, M_{\odot}$. Rapid coolers, C, D, have masses in the interval $M\simeq (1.75 - 2)\, M_{\odot}$. The dependence of the cooling curves on the NS mass remains very  regular for $1\,M_{\odot} < M < 2.149 \, M_\odot$.

In the MKVORH$\phi$ model  the T proton gap  vanishes for $n>3.254\, n_0$.
The central density $n_{\rm cen}=3.254\, n_0$ is reached  for  $M =  1.830 \, M_{\odot}$. The  hyperons do not exist in the objects 8, 4, 5, A, which are slowest    coolers in this model. The objects 1, 2
and intermediate coolers 6, 7, 3, E, Cas A, B and E are described in a narrow mass interval $1.429 \, M_{\odot}<M<1.45 \, M_{\odot}$. Cas A object is described as having the mass $M=(1.45-1.46)\, M_{\odot}$, the object C is described by $M\simeq (1.47-1.5) \, M_{\odot}$ and D, as having the mass in the interval $1.5\, M_{\odot}<M<(1.6-2) \, M_{\odot}$.

In Fig. \ref{NoHhyperon7} (left and right)  we show the same as in Fig. \ref{NoHhyperon3}, \ref{NoHhyperon4}, \ref{NoHhyperon5}, \ref{NoHhyperon6}, but for proton pairing gaps following the   EEHO model.    As we see, the NS in Cas A in the MKVOR model  is appropriately explained by the object with $M\simeq 1.75 \, M_{\odot}$. The dependence of the cooling curves on the NS mass remains  regular for $M_{\odot}<M<2.149 \, M_{\odot}$.

In the MKVORH$\phi$ model   the EEHO proton gap vanishes for $n>3.255\, n_0$, similar to the  T model. The central density $n_{\rm cen}=3.255\, n_0$ is reached  for  $M =  1.830 \, M_{\odot}$. At large densities the proton gap in EEHO model is bigger than that in T model. The  hyperons do not exist in the objects 8, 4, 5, which are most slow    coolers in this model. The intermediate coolers 1, 2, A
and  6, 7, 3, E, Cas A, B and E, are described in the mass interval $1.429\, M_{\odot}<M<1.5\, M_{\odot}$. Cas A object is described as having the mass $M\simeq 1.47 \, M_{\odot}$, the object C is described by $M\simeq (1.6-1.7)\, M_{\odot}$ and D, as having the mass $1.7\, M_{\odot}<M<1.9\, M_{\odot}$.

Comparing cooling curves obtained with various proton gaps, in models BCLL, CCYms, EEHOr, T, EEHO,   we conclude that the cooling picture is more sensitive to the density dependence of the proton gaps than to their values.

In Fig. \ref{NoHhyperon8} (left and right)  we show the same as in Fig. \ref{NoHhyperon3}, \ref{NoHhyperon4}, \ref{NoHhyperon5}, \ref{NoHhyperon6}, \ref{NoHhyperon7},  but for proton pairing gaps following the   CCDK model.  The proton gaps in CCDK model get the highest values compared to those in other models, which we have considered.  As we see in the left panel, all slow coolers in this model (8, 4, 5, A, 1, 2) are described by NSs with masses $1\,M_{\odot}<M<1.8\, M_{\odot}$. Objects 3, 6, 7 are intermediate coolers.  Rapid coolers, B, E, Cas A, C, D, are described by objects having masses $1.9\, M_{\odot}<M<2 \, M_{\odot}$. The NS in Cas A in MKVOR model  is appropriately explained by the high-mass object, with $M\simeq (1.97-2)\, M_{\odot}$.
The dependence of the cooling curves on the NS mass is  weak for $M_{\odot}<M<(1.7-1.8)\, M_{\odot}$ and becomes a more sharp for larger $M$.  For $M < 1.7 \, M_{\odot}$ the effective pion gap (see solid curve in Fig. \ref{piongap}) did not yet reach its minimum value and the MMU emissivity is not as strong thereby. The cooling of massive stars with $M\gsim (1.7-1.8) \, M_{\odot}$ is determined mainly by the MMU processes with a low effective pion gap, $\omega^{*2}(n\geq 3n_0)\simeq 0.3m_{\pi}^2$, see Fig. \ref{piongap}.

In the MKVORH$\phi$ model   the proton gap  vanishes for $n>3.679 \, n_0$. The  central density $n_{\rm cen}=3.679 \, n_0$ is reached  for  $M = 1.990 \, M_{\odot}$. The  hyperons do not exist only in the objects 8,4,5, which are most slow    coolers in this model. In other objects there exist hyperons but, since the CCDK proton gap remains very high, the cooling picture looks more regular than that we had using other proton gaps.
The objects A, 1, 2, 6 are described in the mass interval $(1.45 - 1.5)\, M_{\odot}$. The Cas A object has the mass $M\simeq 1.65\, M_{\odot}$. Objects B, E, C have masses $M\simeq (1.75-1.9)\, M_{\odot}$ and the object D  has the mass $1.9\, M_{\odot}<M<2\, M_{\odot}$. Thus with the CCDK proton gaps the cooling picture looks very regular  in the model including hyperons.

\begin{figure}
	\centerline{\includegraphics[height=7.1cm,clip=true]{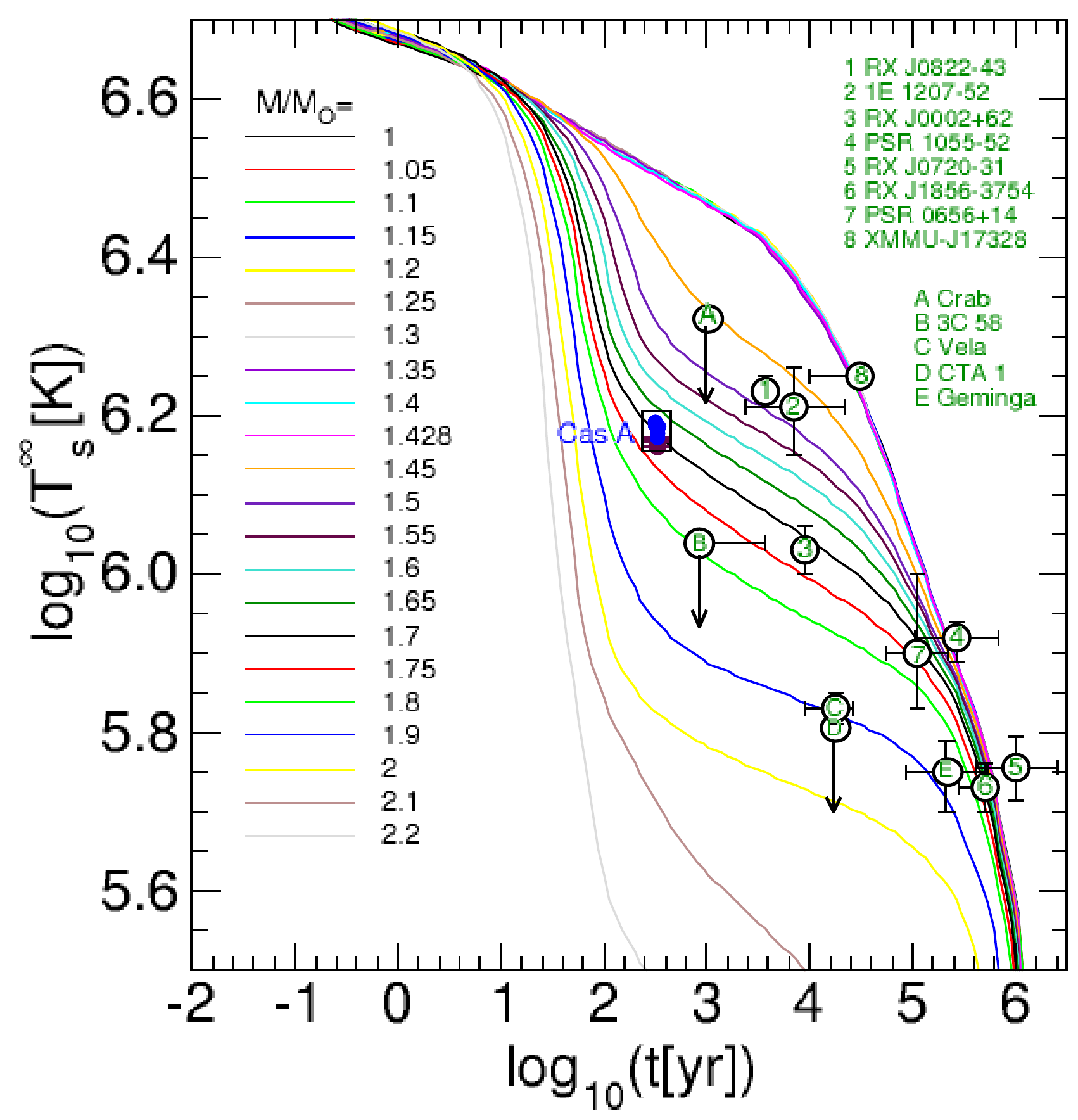}\quad
		\includegraphics[height=7.1cm,clip=true]{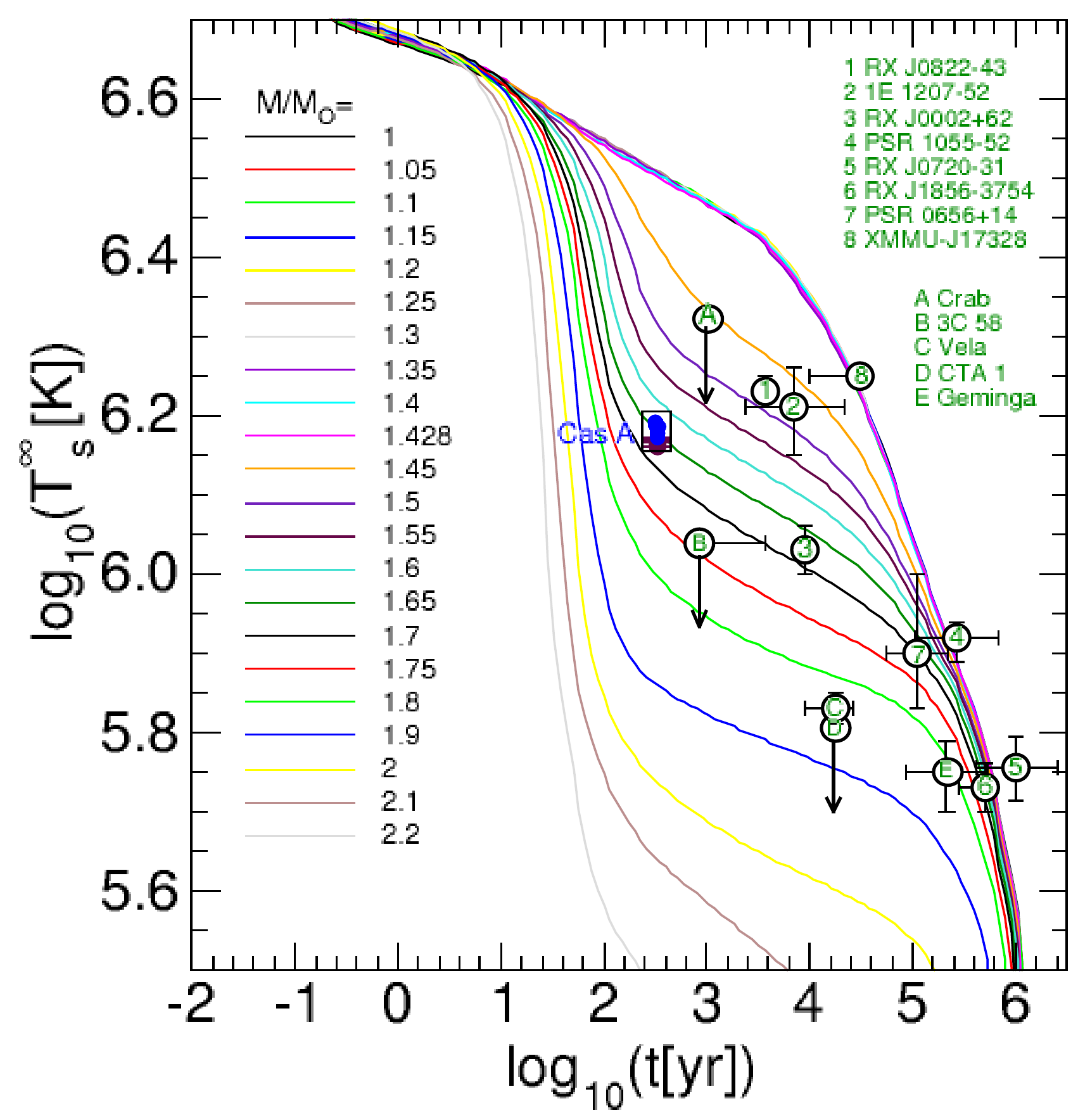}}
	\caption{ Redshifted surface temperature as a function of the NS age for various NS masses for MKVORH model (with hyperons). Proton gaps  are taken following CCDK. Effective pion gap follows the solid curve in Fig. \ref{piongap}.
		Left panel:    Hyperon gaps follow TT1 model.  Right panel:  Hyperon gaps follow TN-NDSoft model.
		model.}
	\label{hyperon9}
\end{figure}

In figures shown above we used TN-FGA model for the $\Lambda$ pairing gaps.
With the hyperon gaps given by TT1, TTGm, TN-NDSoft, TN-Ehime models the cooling curves look very similar to those calculated with TN-FGA parameter choice. As example, in Fig. \ref{hyperon9} we show results obtained  in MKVORH$\phi$ model with the  CCDK proton gaps
and TT1 $\Lambda$ gaps on left panel and for TN-NDSoft $\Lambda$ gaps on right panel. As we see from comparison of the cooling curves in Fig. \ref{hyperon9} (left and right) and Fig. \ref{NoHhyperon8} (right) the dependence on the model for the chosen values of the hyperon gap remains rather weak.

\begin{figure}
	\centerline{
		\includegraphics[height=7.1cm,clip=true]{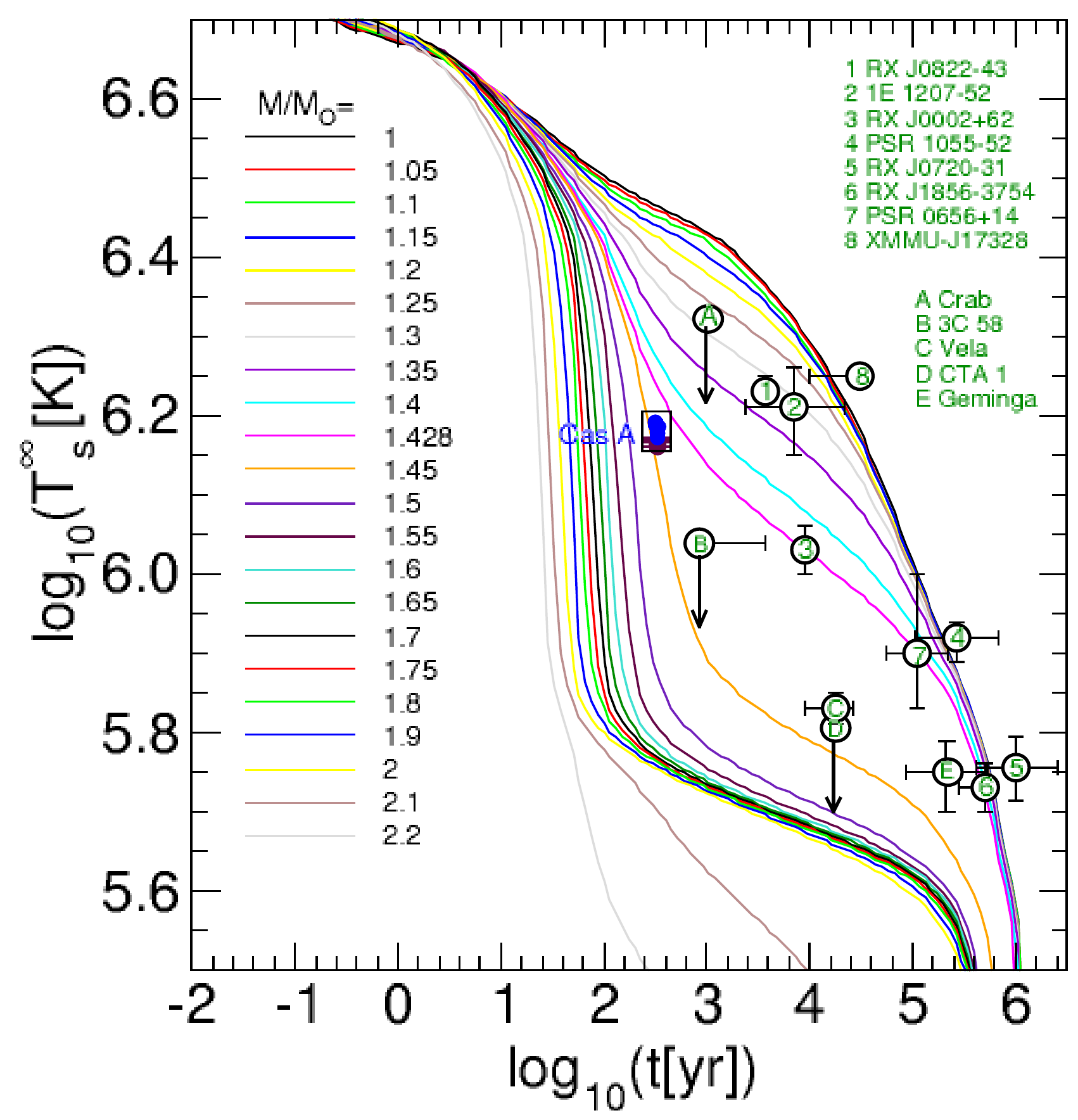}\quad
		\includegraphics[height=7.1cm,clip=true]{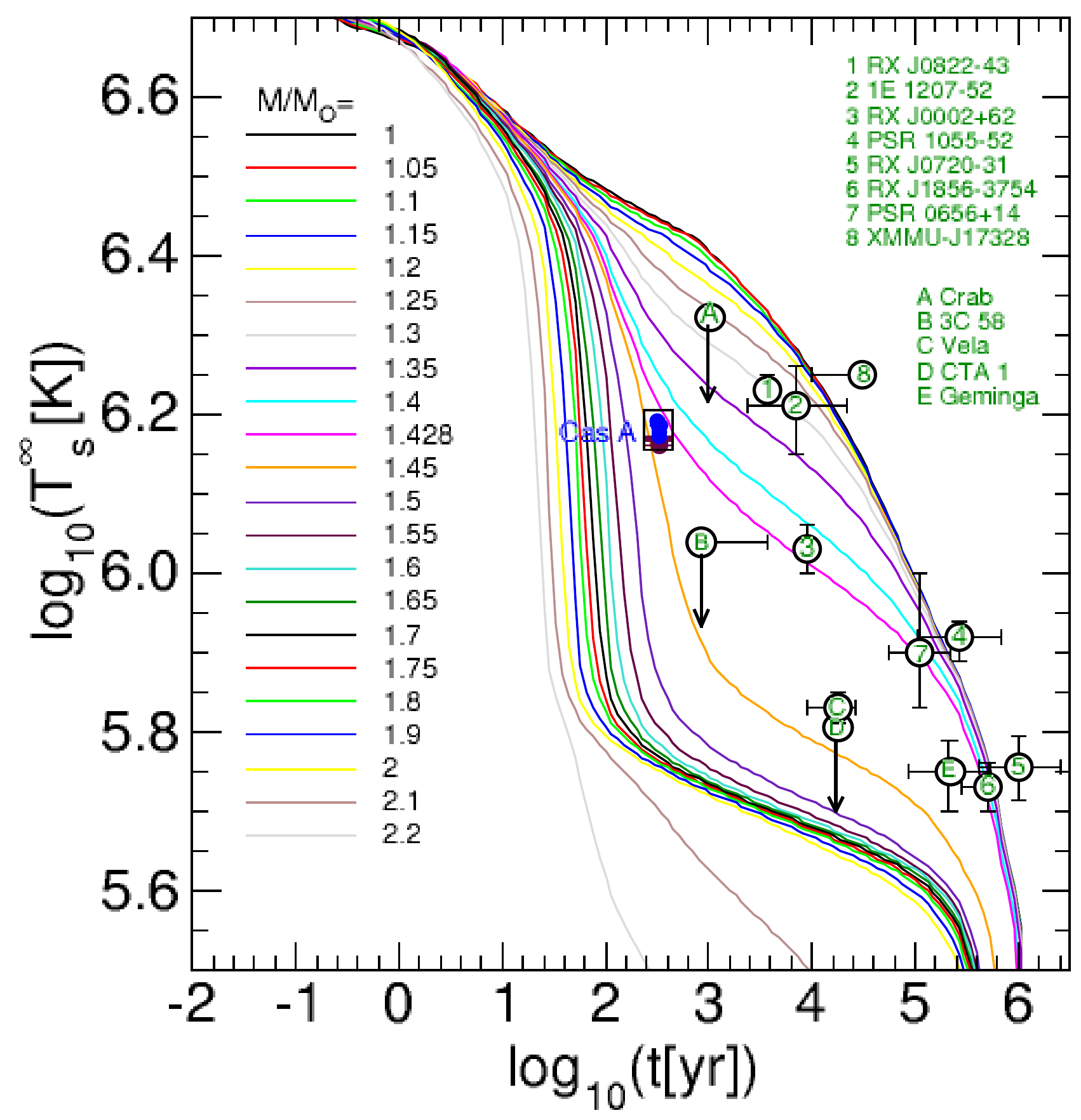}}
	\caption{ Redshifted surface temperature as a function of the NS age for various NS masses for MKVORH model (with hyperons). Proton gaps  are taken following EEHOr model. Hyperon gaps follow TN-FGA model.
		Left panel:   Effective pion gap follows the dotted curve in Fig. \ref{piongap}. Right panel:  Effective pion gap follows the dash-dotted   curve in Fig. \ref{piongap}.
		model.}
	\label{hyperon4}
\end{figure}

In Fig. \ref{hyperon4} we demonstrate a dependence of the cooling picture on the choice of the effective pion gap. For that on the left panel we show the cooling history for the MKVORH$\phi$ model with the EEHOr  proton gaps  and the effective pion gap given by the dotted curve  in Fig. \ref{piongap} (for $n_c^{\rm PU}=2\, n_0$) and on the right panel, for the effective pion gap  given by the dash-dotted curve for $n_c^{\rm PU}=1.5\, n_0$. Comparing the cooling pictures in Fig. \ref{hyperon4} (left and right) and   on right panel of Fig. \ref{NoHhyperon5} we see that a dependence is visible for $M<M_{c,\Lambda}^{\rm DU} \simeq 1.429\, M_{\odot}$, when the main cooling regulator is the MMU process. The slow coolers, such as XMMU, are better described with the effective pion gap following the solid curve in Fig. \ref{piongap}. For $M>1.429\, M_{\odot}$ the main cooling regulator is already the DU process on $\Lambda$s and thereby
the cooling curves become not sensitive to the values of the effective pion gap.

\section{Conclusion}
\label{sect::conclusion}
In the given work we exploited the equation of state of the hadronic MKVOR model (assuming absence of hyperons) and the MKVORH$\phi$ model (with inclusion of hyperons). We have demonstrated that the presently known cooling data  can be
appropriately described within our nuclear medium cooling scenario both without inclusion of hyperons (MKVOR model) and with included  hyperons (within MKVORH$\phi$ model) under assumption that different sources have different masses, provided  we use appropriately selected models for proton gaps, e.g. following the models BCLL,  CCYms, EEHOr, T, EEHO, CCDK, see left panel of  Fig. \ref{Protongaps}, and  the models TT1, TTGm, TN-NDSoft, TN-Ehime, TN-FGA for $\Lambda$ pairing gaps, see right panel of Fig. \ref{Protongaps}.  To larger densities spreads the proton gap and the higher values it  gets within the density interval of its existence, the more regular is  the neutron-star mass dependence of the cooling curves in our scenario.


The effect of hyperons is rather strong in MKVORH$\phi$ model,   after the DU processes with  participation of $\Lambda$s are allowed (for $M>1.429\, M_{\odot}$). The so called ``strong DU constraint" of \cite{Klahn:2006ir} that the DU processes  should not be allowed in neutron stars with $M< 1.5 \, M_{\odot}$ is not fulfilled for the DU processes with hyperons. However one should bear in mind that the DU emissivity on hyperons is typically two orders of magnitude smaller than that for the DU processes on nucleons. Therefore the strong DU constraint should be partially relaxed.
Besides, note that  influence of hyperons on the neutron-star cooling would be much less significant, if we used the KVORcut-based models to construct the equation of state, cf. \cite{Maslov:2015wba}.


Besides hyperons, another contribution to the neutrino emissivity could come from the processes with $\Delta$-isobars. Among them the density for appearance of $\Delta^-$ is the lowest because it becomes possible to replace the leptons by $\Delta^-$ to fulfill the charge neutrality condition \cite{Kolomeitsev:2016ptu, Li:2018qaw}. However, the threshold baryon density for the DU processes on $\Delta^-$ is
larger than the one for the nucleon DU processes, since the $\Delta^-$ fraction is always lower than the proton one \cite{Prakash:1992zng}. Within the MKVOR-based model used in the given work appearance of the $\Delta$s leads to a notable increase of the proton fraction, so the critical density for the nucleon DU processes decreases. Nevertheless  the critical value of the  neutron-star mass $M^{\rm DU}_{c, p}$ remains higher than 1.5~$M_\odot$, provided  the $\Delta$ potential  satisfies inequality $U_\Delta (n_0)> -88$~MeV, whereas a realistic value of the $\Delta$ potential  is $U_\Delta (n_0)\sim  -50$~MeV. On the other hand the hyperon fractions change only a little after  inclusion of $\Delta$s \cite{Kolomeitsev:2016ptu}. Thus in this work devoted to the study of the neutron-star cooling in hadronic models with hyperons we disregarded the possibility of appearance of $\Delta$s. In a more detail the latter possibility will be studied elsewhere.

We focused on the study of purely hadronic equation of state. Another possibility is to consider cooling of hybrid stars within this equation of state  for the hadron phase  and allowed the first order phase transition from the hadron to the quark matter, cf. \cite{Blaschke:2000dy,Grigorian:2004jq,Sedrakian:2015qxa}, and permitted various  mixed phases, cf.  \cite{Voskresensky:2002hu,Maruyama:2005tb,Maruyama:2007ey,Ayriyan:2017nby}  and refs therein. We hope to return to such a study in subsequent works.

\section*{Acknowledgements}
We thank D. Blaschke and E. E. Kolomeitsev for valuable discussions
and suggestions.  The research  was supported    by the Ministry of Education and Science of the Russian Federation within the state assignment,  project No 3.6062.2017/6.7. We acknowledge as well the support of the Russian Science Foundation, project No 17-12-01427.


\end{document}